\documentclass[journal]{IEEEtran}
\usepackage{amsmath,amssymb,graphicx,cases}
\usepackage{url,hyperref}
\usepackage{amsfonts,enumitem}
\usepackage[english]{babel}
\usepackage{algorithm,bm}
\usepackage{mdframed}
\usepackage[utf8]{inputenc}
\usepackage[noend]{algpseudocode}
\usepackage{comment}
\ifCLASSOPTIONcompsoc
  \usepackage[caption=true,font=footnotesize,labelfont=sf,textfont=sf]{subfig}
\else
  \usepackage[caption=false,font=footnotesize]{subfig}
\fi

\def\algbackskip{\hskip-\ALG@thistlm}
\newtheorem{theo}{Theorem}
\newtheorem{lemma}{Lemma}
\def \bF{\boldsymbol{F}}
\def \bE{\boldsymbol{E}}
\def \bG{\boldsymbol{G}}
\def \bI{\boldsymbol{I}}
\begin{document}

\def \bx {\boldsymbol{x}}
\def \by {\boldsymbol{y}}
\def \bu {\boldsymbol{u}}
\def \bz {\boldsymbol{z}}
\def \bzero {\boldsymbol{0}}
\def \bdelta {\boldsymbol{\delta}}

\newcommand*{\QEDA}{\hfill\ensuremath{\clubsuit}}%

%
\title{Signal Recovery in Perturbed \\ Fourier Compressed Sensing}
\author{Eeshan~Malhotra,~\IEEEmembership{Student Member,~IEEE,} 
		Himanshu~Pandotra,~\IEEEmembership{}
        Ajit~Rajwade,~\IEEEmembership{Member,~IEEE} and
        Karthik S. Gurumoorthy~\IEEEmembership{}
\thanks{Eeshan Malhotra and Himanshu Pandotra are both first authors with equal contribution. Eeshan Malhotra and Ajit Rajwade are with the Department of Computer Science and Engineering at IIT Bombay. Himanshu Pandotra is with the Department of Electrical Engineering at IIT Bombay. Karthik Gurumoorthy is with the International Center for Theoretical Sciences, Bengaluru.  The email addresses of the authors are \url{eeshan@gmail.com}, \url{angad@ee.iitb.ac.in}, \url{ajitvr@cse.iitb.ac.in} and \url{karthik.gurumoorthy@icts.res.in} respectively. Corresponding author is Ajit Rajwade. Ajit Rajwade acknowledges generous support from IITB seed grant \#14IRCCSG012. Karthik Gurumoorthy thanks the AIRBUS Group Corporate Foundation Chair in Mathematics of Complex Systems established in ICTS-TIFR.}}

\markboth{IEEE Transactions on Signal Processing}%
{\ \ }



\maketitle

\begin{abstract}
In many applications in compressed sensing, the measurement matrix is a Fourier matrix, \textit{i.e.}, it measures the Fourier transform of the underlying signal at some specified `base' frequencies $\{u_i\}_{i=1}^M$, where $M$ is the number of measurements. However due to system calibration errors, the system may measure the Fourier transform at frequencies $\{u_i + \delta_i\}_{i=1}^M$ that are different from the base frequencies and where $\{\delta_i\}_{i=1}^M$ are unknown. Ignoring perturbations of this nature can lead to major errors in signal recovery. In this paper, we present a simple but effective alternating minimization algorithm to recover the perturbations in the frequencies \emph{in situ} with the signal, which we assume is sparse or compressible in some known basis. In many cases, the perturbations $\{\delta_i\}_{i=1}^M$ can be expressed in terms of a small number of unique parameters $P \ll M$. We demonstrate that in such cases, the method leads to excellent quality results that are several times better than baseline algorithms (which are based on existing off-grid methods in the recent literature on direction of arrival (DOA) estimation, modified to suit the computational problem in this paper). Our results are also robust to noise in the measurement values. We also provide theoretical results for (1) the convergence of our algorithm, and (2) the uniqueness of its solution under some restrictions.
\end{abstract}


%
\IEEEpeerreviewmaketitle

\section{Introduction}
%
%
%
%
\IEEEPARstart{C}{ompressed} sensing (CS) is today a very widely researched branch of signal and image processing. Consider a vector of compressive measurements $\boldsymbol{y} \in \mathbb{C}^M, \boldsymbol{y} = \boldsymbol{\Phi x}$ for signal $\boldsymbol{x} \in \mathbb{C}^N$, acquired through a sensing matrix $\boldsymbol{\Phi} \in \mathbb{C}^{M \times N}, M < N$. CS theory offers guarantees on the error of reconstruction of $\boldsymbol{x}$ that is sparse or compressible in a given orthonormal basis $\boldsymbol{\Psi} \in \mathbb{C}^{N \times N}$, assuming that the sensing matrix (also called measurement matrix) $\boldsymbol{\Phi} \in \mathbb{C}^{M \times N}$ (and hence the product matrix $\boldsymbol{\Phi \Psi}$) obeys some properties such as the restricted isometry (RIP) \cite{Candes2008}. Moreover, the guarantees apply to efficient algorithms such as basis pursuit. However the underlying assumption is that the sensing matrix $\boldsymbol{\Phi}$ is known accurately. If $\boldsymbol{\Phi}$ is known inaccurately, then signal-dependent noise will be introduced in the system causing substantial loss in reconstruction accuracy. 

Of particular interest in many imaging applications such as magnetic resonance imaging (MRI), tomography or Fourier optics \cite{Katz2010,Schniter2015,Lustig2008,Ferrucci2015}, is the case where the measurement matrix is a row-subsampled version of the Fourier matrix, where the frequencies may or may not lie on a Cartesian grid of frequencies used in defining the Discrete Fourier Transform (DFT). However, it is well-known that such Fourier measurements are prone to inaccuracies in the acquisition frequencies. This may be due to an imperfectly calibrated sensor. In case of specific applications such as MRI, this is due to perturbations introduced by gradient delays in the MRI machine \cite{gd1jang2016rapid,gd2robison2010fast,Brodsky2009}. In case of computed tomography (CT), it may be due to errors in specification of the angles of tomographic acquisition due to geometric calibration errors in a CT machine \cite{Ferrucci2015}, or in the problem of tomographic under unknown angles \cite{Malhotra2016}. 

\subsection{Relation to Previous Work}
The problem we deal with in this paper is a special case of the problem of `blind calibration' (also termed `self-calibration') where perturbations in the sensing matrix are estimated \textit{in situ} along with the signal. Here, we expressly deal with the case of Fourier sensing matrices with imperfectly known frequencies. There exists a decent-sized body of earlier literature on the general blind calibration problem (not applied to Fourier matrices) beginning with theoretical bounds derived in \cite{Herman2010}. Further on, \cite{zhang2012robustly} analyze a structured perturbation model of the form $\boldsymbol{y} = (\boldsymbol{A} + \boldsymbol{B \Delta})\boldsymbol{x}$ where $\boldsymbol{x},\boldsymbol{\Delta}$ are the unknown signal and diagonal matrix of perturbation values respectively, and $\boldsymbol{A},\boldsymbol{B}$ are the fully known original sensing matrix and perturbation matrix respectively. The theory is then applied to direction of arrival (DOA) estimation in signal processing. Further work in \cite{nehorai2014structured} uses the notion of group-sparsity to infer the signal $\boldsymbol{x}$ and the perturbations $\boldsymbol{\Delta}$ using a convex program based on a first order Taylor expansion of the parametric DOA matrix. A total least squares framework that also accounts for sparsity of the signal is explored in \cite{Zhu2011} for a perturbation model of the form $\boldsymbol{y} + \boldsymbol{e} = (\boldsymbol{A}+\boldsymbol{E})\boldsymbol{x}$ where $\boldsymbol{e},\boldsymbol{E}$ are the additive errors in the measurement vector $\boldsymbol{y}$ and measurement matrix $\boldsymbol{A}$ respectively. In \cite{Ling2016}, \cite{Ling2015}, \cite{Bilen2014},\cite{Cambareri2016}, the following framework is considered: $\boldsymbol{y} = \boldsymbol{\Delta A x}$, where $\boldsymbol{\Delta}$ is a diagonal matrix containing the unknown sensor gains which may be complex, $\boldsymbol{x}$ is the unknown sparse signal, and $\boldsymbol{A}$ is the known sensing matrix. Both $\boldsymbol{x}$ and $\boldsymbol{\Delta}$ are recovered together via linear least squares in \cite{Ling2016}, via the lifting technique on a biconvex problem in \cite{Ling2015}, using a variety of convex optimization tools in \cite{Bilen2014}, and in \cite{Cambareri2016} using a non-convex method. The problem we deal with in this paper cannot be framed as a single (per measurement) unknown phase or amplitude shift/gain unlike these techniques, and hence is considerably different. 

Related to (but still very different from) the aforementioned problem of a perturbed sensing matrix, is the problem of a perturbed or mismatched signal representation matrix $\boldsymbol{\Psi}$ which can also cause significant errors in compressive recovery \cite{Chi2011}. This has been explored via alternating minimization in \cite{Nichols2014}, via a perturbed form of orthogonal matching pursuit (OMP) in \cite{Teke2013}, and via group-sparsity in \cite{nehorai2014structured}. The problem of estimating a small number of complex sinusoids with off-the-grid frequencies from a subset of regularly spaced samples has been explored in \cite{Tang2013}. Note that in \cite{Chi2011,Tang2013,nehorai2014structured,Nichols2014}, the emphasis is on mismatch in the representation matrix $\boldsymbol{\Psi}$ and \emph{not} in the sensing matrix $\boldsymbol{\Phi}$ - see Section \ref{sec:comp_basismismatch} for more details. The problem of \emph{additive} perturbations in both the sensing matrix as well as the representation matrix has been analyzed in \cite{perturbaldroubi2012perturbations}, using several assumptions on both perturbations. Note that the perturbations in the Fourier sensing matrix do not possess such an additive nature.

To the best of our knowledge, there is no previous work on the analysis of perturbations in a Fourier \emph{measurement} matrix in a \emph{compressive sensing} framework. Some attempts have been made to account for frequency specification errors in MRI, however, most of these require a separate off-line calibration step where the perturbations are measured. However in practice, the perturbations in frequencies may be common to only subsets of measurements (or even vary with each measurement), and need not be static. In cases where the correction is made alongside the recovery step, a large number of measurements may be required \cite{mri1ianni2016trajectory}, as the signal reconstruction does not deal with a compressed sensing framework involving $\ell_q$ ($q < 1$) minimization. The problem of perturbations in the Fourier matrix also occurs in computed tomography (CT). This happens in an indirect way via the Fourier slice theorem, since the 1D Fourier transform of a parallel beam tomographic projection in some acquisition angle $\alpha$ is known to be equal to a slice of the Fourier transform of the underlying 2D image at angle $\alpha$. In CT, the angles for tomographic projection may be incorrectly known due to geometric errors \cite{Ferrucci2015} or subject motion, and uncertainty in the angle will manifest as inaccuracy of the Fourier measurements. Especially in case of subject motion, the measurement matrix will contain inaccuracies that cannot be pre-determined, and must be estimated \textit{in situ} along with the signal. While there exist approaches to determine even the completely unknown angles of projection, they require a large number of angles, and also the knowledge of the distribution of the angles \cite{tomo1,tomo2}. Our group  has presented a method \cite{Malhotra2016} which does not require this knowledge, but in \cite{Malhotra2016}, the angles are estimated only along with the image moments. The image itself is estimated after determining the angles. In contrast, in this paper, the errors in frequency are determined along with the underlying signal.

A large body of existing work is also lacking in theoretical backing. For instance \cite{perturbaldroubi2012perturbations} makes assumptions on the properties of the perturbed measurement matrix, such as the magnitudes of the perturbations. Some existing approaches to handle perturbations in $\boldsymbol{\Psi}$ simplify the problem using a Taylor approximation \cite{nehorai2014structured,zhang2012robustly,Fannjiang2013}. However, when such an approach is tailored to the problem of perturbation in $\boldsymbol{\Phi}$, it proves to be adequate only at extremely small perturbation levels in our case, rendering the adjustment for the perturbation to be much less effective (See Section \ref{sec:results}).

\textbf{Contributions:} A method for simultaneous recovery of the perturbations and the signal in a perturbed Fourier compressed sensing structure is proposed in this paper. The algorithm is verified empirically over a large range of simulated data under noise-free and noisy cases. Further, we analyze the convergence of the algorithm, as well as the uniqueness of the solution to our specific computational problem under specific but realistic assumptions about the measurement perturbations. We also provide guarantees on the recovered signal given a linearized approximation of the original objective function, and also analyze the reconstruction error for an average sensing matrix if the perturbations in the Fourier measurements were ignored. 

\subsection{Organization of the Paper}
This paper is organized as follows. Section \ref{sec:probdef} defines the problem statement. The recovery algorithm is presented in Section \ref{sec:alg}, followed by extensive numerical results in Section \ref{sec:results}. The theoretical treatment is covered in Section \ref{sec:theory}, followed by a conclusion in Section \ref{sec:conclusion}

\section{Problem Definition}
\label{sec:probdef}
Formally, let $\boldsymbol{F} \in \mathbb{C}^{M\times N}$ be a Fourier matrix using a known (possibly, but not necessarily on-grid) frequency set $\boldsymbol{u} \triangleq \{u_i\}_{i=1}^M \in \mathbb{R}^M$, $\boldsymbol{x} \in \mathbb{R}^N$  be a signal that is sparse (with at the most  $s$ non-zero values) or compressible, measured using a perturbed Fourier matrix $\boldsymbol{F_{t}} \in \mathbb{C}^{M \times N}$. That is, 
\begin{equation}
    \boldsymbol{y} = \boldsymbol{F_{t}x} + \boldsymbol{\eta},
    \label{eq:eq1}
\end{equation}
where, $\boldsymbol{\eta}$ is a signal-independent noise vector, $\boldsymbol{F_{t}}$ is a Fourier measurement matrix at the set of unknown frequencies $\boldsymbol{u} + \boldsymbol{\delta} \triangleq \{u_i + \delta_i\}_{i=1}^M$, with $\forall i, \delta_i \in \mathbb{R},|\delta_i| \leq r, r \geq 0,  \boldsymbol{\delta} \triangleq \{\delta_i\}_{i=1}^M$. Note that we assume full knowledge of $\{u_i\}_{i=1}^M$, \textit{i.e.}, the base frequencies. The problem is to recover \textit{both}, the sparse signal $\boldsymbol{x}$, and the unknown perturbations in the frequencies, $\boldsymbol{\delta}$. This is formalized as the following:
\begin{equation}
\underset{\boldsymbol{\hat{x}}, \boldsymbol{\hat{\delta}} \in [-r,r]^M}{\text{min}} J(\boldsymbol{\hat{x}},\boldsymbol{\hat{\delta}}) \triangleq \|\boldsymbol{\hat{x}}\|_1 + \lambda\|\boldsymbol{y} - \boldsymbol{\hat{F}}(\boldsymbol{\hat{\delta}})\boldsymbol{\hat{x}} \|_2 
\label{eq:gen_prob}
\end{equation}
where $\boldsymbol{\hat{F}}(\boldsymbol{\hat{\delta}})$ is the Fourier measurement matrix at frequencies $\boldsymbol{u + \hat{\delta}}$, and $\boldsymbol{\hat{\delta}}$ denotes the estimate of $\boldsymbol{\delta}$. Note that the above problem is a perturbed version of the so-called square-root LASSO (SQ-LASSO), since the second term involves an $\ell_2$ norm and not its square. We used the SQ-LASSO due to its advantages over the LASSO in terms of parameter tuning, as mentioned in \cite{Belloni2011}. 

Eqn. \ref{eq:gen_prob} presents the most general formulation of the problem. The signal may be sparse in a non-canonical basis, say the Discrete Wavelet transform (DWT), in which case the objective function in Eqn. \ref{eq:gen_prob} can be changed, leading to the following problem:
\begin{equation}
\underset{\boldsymbol{\hat{\theta}}, \boldsymbol{\hat{\delta}} \in [-r,r]^M}{\text{min}} J(\boldsymbol{\hat{\theta}},\boldsymbol{\hat{\delta}}) \triangleq \|\boldsymbol{\hat{\theta}}\|_1 + \lambda\|\boldsymbol{y} - \boldsymbol{\hat{F}}(\boldsymbol{\hat{\delta}}) \boldsymbol{\Psi \hat{\theta}} \|_2, 
\end{equation}
where $\boldsymbol{\theta} = \boldsymbol{\Psi}^T \boldsymbol{x}$ are the wavelet coefficients of $\boldsymbol{x}$. We also discuss an important modification. In Eqn. \ref{eq:gen_prob}, we have assumed that all perturbations, i.e. entries in $\boldsymbol{\delta}$ are independent. However, this may not necessarily be the case in many applications. For example, consider the following three cases (though the applicability of our technique and analysis is not restricted to just these):
\begin{enumerate}
\item Consider parallel beam tomographic reconstruction of a 2D signal $f(x,y)$ with incorrectly specified angles. The 1D Fourier transform of the tomographic projection of $f$ acquired at some angle $\alpha$ is equal to a slice through the 2D Fourier transform of $f$ at angle $\alpha$ and passing through the origin of the Fourier plane. The frequencies along this slice can be expressed in the form $u^{(1)} = \rho \cos \alpha, u^{(2)} = \rho \sin \alpha$ where $\rho = \sqrt{(u^{(1)})^2 + (u^{(2)})^2}$ is the distance between $(u^{(1)},u^{(2)})$ and $(0,0)$ in frequency-space. If the specified angle has an error $\bar{\alpha}$, the effective Fourier measurements are at frequencies $\bar{u}^{(1)} = \rho \cos (\alpha + \bar{\alpha}), \bar{u}^{(2)} = \rho \sin (\alpha + \bar{\alpha})$. In such a case, the perturbations in all the frequencies along a single slice are governed by a \emph{single} parameter $\bar{\alpha}$ which is unknown. (The parameter $\rho$ is known since the base frequencies $(u^{(1)},u^{(2)})$ are known.) This basic principle also extends to other projection methods such as cone-beam and to higher dimensions. (See Fig. \ref{fig:fig7} for sample reconstructions for this application).
\item The problem of tomography under unknown angles is of interest in cryo-electron microscopy to determine the structure of virus particles \cite{Luvcic2013}. Here the angles of tomographic projection as well as the underlying image are both unknown. In some techniques, the angles of projection are estimated first using techniques from dimensionality reduction \cite{tomo2} or geometric relationships \cite{Malhotra2016,tomo1}. Any error in the angle estimates affects the estimate of the underlying image in a manner similar to that described in the previous point. 
\item In MRI, gradient delays can cause errors in the specified set of frequencies at which the Fourier transform is measured \cite{Brodsky2009}. The gradient delays are essentially the difference between the programmed or specified start time and the start time which the machine uses for the measurement. For a single axis, the gradient $G(t)$ would produce a trajectory of measurements of the form $k(t) = K \int_{\tau = 0}^t G(\tau) d\tau$ at time $t$ where $K$ is a hardware-related proportionality constant and $u(t) \triangleq (u^{(1)}(t),u^{(2)}(t)$ for 2D measurements. Given a gradient delay of $\bar{t}$, the actual trajectory would be $k'(t) = K \int_{\tau = 0}^t G(\tau - \bar{t}) d\tau$. For small-valued $\bar{t}$, this leads to a trajectory error proportional to $G(t) \bar{t}$ \cite{Luvcic2013}. Thus frequency perturbations in MRI measurements for a single axis are governed by a single parameter $\bar{t}$. In some specific MRI sampling schemes such as radial, a single global trajectory error is assumed for all frequencies in one or all radial spokes (see Eqn. 3 of \cite{Moussavi2014}, and `Methods section' in \cite{Deshmane2016}). This global error arises due to gradient delays, which again presents a case of perturbations in multiple measurements being expressed in terms of a single parameter.
\end{enumerate}
Handling cases such as these in fact makes the recovery problem more tractable, since the number of unknowns is essentially reduced. We now present our recovery algorithm and its modified version for handling cases where many measurements share a common set of `perturbation parameters', in the following section. The convergence of the algorithm is analyzed in Section \ref{sec:convergence}.

\section{Recovery Algorithm}
\label{sec:alg}
We present an algorithm to determine $\boldsymbol{x}$ and $\boldsymbol{\delta}$ by using an alternation between two sub-problems. Starting with a guess $\boldsymbol{\hat{\delta}}$ for the perturbations $\boldsymbol{\delta}$, we recover $\boldsymbol{\hat{x}}$, an estimate for $\boldsymbol{x}$, using the SQ-LASSO mentioned before, which is essentially an unconstrained $l_1$ norm minimization approach common in compressive sensing. Next, using this first estimate $\boldsymbol{\hat{x}}$, we update $\boldsymbol{\hat{\delta}}$ to be the best estimate, assuming $\boldsymbol{\hat{x}}$ to be the truth, using a linear brute force search in the range $-r$ to $r$. A linear search is possible because each measurement $y_i$ is the dot product of a single row of $\boldsymbol{F_t}$ with $\boldsymbol{x}$, and hence a single $(u_i, \delta_i)$ value is involved. Consequently, the different $\delta_i$ values can be recovered through \emph{independent} parallel searches (see Section \ref{sec:comp_basismismatch} for a comparison to related computational problems). From here on, we alternate between the two steps - recovery of $\boldsymbol{\hat{x}}$ and recovery of $\boldsymbol{\hat{\delta}}$, till convergence is achieved. 

Since the search space is highly non-convex, we also employ a multi-start strategy, where, we perform multiple runs of the alternating algorithm to recover $\boldsymbol{\hat{\delta}}$ and $\boldsymbol{\hat{x}}$, each time, initializing the first guess for $\boldsymbol{\hat{\delta}}$ randomly. We ultimately select the  solution that minimizes the objective function $J(\boldsymbol{\hat{x}},\boldsymbol{\hat{\delta}})$. In practice, we have observed that the number of starts required for a good quality solution is rather small (around 10). 

The full algorithm, including the optimization for multi-start is presented in Algorithm \ref{A1}. Note that $\boldsymbol{\hat{F}_k}({\hat{\delta_k}})$ denotes the $k^{\textrm{th}}$ row of $\boldsymbol{\hat{F}}({\boldsymbol{\hat{\delta}}})$.
\begin{algorithm}[!htb]
\caption{Alternating Minimization Algorithm}\label{A1}
\begin{algorithmic}[1]
\Procedure{AlternativeRecovery}{}
\State $converged \gets False, \chi \gets 0.0001$
\State $\boldsymbol{\hat{\delta}} \gets $ sample from $\textrm{Uniform}[-r,+r]$
\While{$converged == False$}
\State $\boldsymbol{\hat{F}} \gets$ Fourier matrix at $(\boldsymbol{u} + \boldsymbol{\hat{\delta}})$
\State Estimate $\boldsymbol{\hat{x}}$ as:
\State $\underset{\boldsymbol{\hat{x}}}{\text{min}} \|\boldsymbol{\hat{x}}\|_1 + \lambda\|\boldsymbol{y} - \boldsymbol{\hat{F}}(\boldsymbol{\hat{\delta}})\boldsymbol{\hat{x}}\|_2 $ 
\State 
\For{$k$ in $1 \to M$} \label{deltafor}
    \State \label{mindelta} Test each discretized value of ${\hat{\delta}}_k$ in range
    \Statex \hspace{18mm}  $-r$ to $r$ and select the value to achieve \\\hspace{18mm}  $\underset{{\hat{\delta}}_k}{\text{min}} \|y_k - \boldsymbol{\hat{F}_k}({\hat{\delta}_k})\boldsymbol{\hat{x}} \|^2 $ 
\EndFor
\If {$\|\boldsymbol{\hat{\delta}} - \boldsymbol{\hat{\delta}}_{prev}\|_2 < \chi$ and
 \item[\hspace{18mm}   $\|\boldsymbol{\hat{x}} - \boldsymbol{\hat{x}}_{prev}\|_2 < \chi$}]
\State $converged \gets True$
\EndIf
\EndWhile
\State \Return $\boldsymbol{\hat{x}}, \boldsymbol{\hat{\delta}}$
\item[]
\EndProcedure

\Procedure{Multistart}{}
\State $\textit{minobjective} \gets \infty$
\State $\boldsymbol{\hat{x}}_{best} \gets null$
\State $\boldsymbol{\hat{\delta}}_{best} \gets null$
\For{start in $1 \to$ numstarts}
    \State $\boldsymbol{\hat{x}}, \boldsymbol{\hat{\delta}} \gets $ AlternatingRecovery()
    \State $\boldsymbol{\hat{F}} \gets$ Fourier matrix at $(u + \boldsymbol{\hat{\delta}})$
    \State $\textit{objective} \gets \|\boldsymbol{\hat{x}}\|_1 + \lambda\|y - \boldsymbol{\hat{F}}\boldsymbol{\hat{x}}(\boldsymbol{\hat{\delta}})\|_2$  
    \If{$\textit{objective} < \textit{minobjective}$}
        \State $\boldsymbol{\hat{x}}_{best} \gets \boldsymbol{\hat{x}}$
        \State $\boldsymbol{\hat{\delta}}_{best} \gets \boldsymbol{\hat{\delta}}$
        \State $\textit{minobjective} \gets \textit{objective}$       
    \EndIf
\EndFor
\State \Return $\boldsymbol{\hat{x}}_{best}, \boldsymbol{\hat{\delta}}_{best}$
\EndProcedure
\end{algorithmic}
\end{algorithm}
We now consider the important and realistic cases where values in $\boldsymbol{\hat{\delta}}$ can be expressed in terms of a small number of unique parameters $\boldsymbol{\beta} \triangleq \{\beta_i\}_{i=1}^P$ where $P \ll M$. We henceforth term these `perturbation parameters'. In other words, there are subsets of measurements whose frequency perturbation values are expressed fully in terms of a \emph{single} perturbation parameter from $\boldsymbol{\beta}$ (besides the base frequency itself). We assume that $\forall k, 1 \leq k \leq P, |\beta_k| \leq r$, where $r > 0$ is known. Let the $k^{\textrm{th}}$ unique value in $\boldsymbol{\beta}$ correspond to the perturbation parameter for measurements in a set $L_k$, indexing into the measurement vector $\boldsymbol{y}$. Thus $\forall i \in L_k, \delta_i = h(\beta_k,u_i)$ where $h$ is a known function of the perturbation parameter $\beta_k$ and base frequency $u_i$. The exact formula for $h$ is dictated by the specific application. 

For example, in the CT application cited at the end of the previous section, let us define set $L_k$ to contain indices of all frequencies along the $k^{\textrm{th}}$ radial spoke at some angle $\alpha_k$. The perturbation values $\delta_i$ for all base frequencies $u_i$ in $L_k$ can be expressed in terms of a single parameter - the error $\beta_k$ in specifying the angle. Here, for frequency $u_i = (u^{(1)}_i,u^{(2)}_i)$, we would have $\delta_i = h(\beta_k,u_i) \triangleq (\rho_i (\cos (\alpha_k + \beta_k) - \cos \beta_k),\rho_i (\sin (\alpha_k + \beta_k) - \sin \alpha_k) )$ where $\rho_i = \sqrt{(u^{(1)}_i)^2 + (u^{(2)}_i)^2}, u^{(1)}_i = \rho_i \cos \alpha_k,  u^{(2)}_i = \rho_i \sin \alpha_k$. In the MRI example, the perturbation values for all base frequencies $u_i$ along the $k^{\textrm{th}}$ axis can be expressed in terms of a single perturbation parameter $\beta_k$, which stands for the gradient delay for the $k^{\textrm{th}}$ axis. In this case, $\delta_i = h(\beta_k,u_i) \triangleq (K'\beta_k G_x(t), K'\beta_k G_y(t)) $ for hardware-related proportionality constant $K'$ and where $G_x(t), G_y(t)$ are the $x,y$ components of the gradient at time $t$ (at which the Fourier transform at frequency $u_i + \delta_i$ was measured). In the case of radial MRI, the parallel and perpendicular components of the error at every frequency in the trajectory along the radial spoke at angle $\alpha$ are expressed as $\delta_{par} = K(t_x \cos^2 \alpha + t_y \sin^2 \alpha),\delta_{perp} = K(-t_x \cos \alpha \sin \alpha + t_y \sin \alpha \cos \alpha)$ where $t_x,t_y$ represent gradient  delays \cite{Deshmane2016} and $K$ is a hardware-related constant. Here, the perturbation parameters are $\beta_1 = t_x, \beta_2 = t_y$, and they are common to all radial spokes.

To suit these cases of perturbation parameters common to many measurements, we modify Algorithm \ref{A1}, for which Step \ref{deltafor} can then be replaced by:
\begin{align*}
&\textbf{for } k \textrm{ in } 1 \rightarrow P \textbf{ do}\\
& \hspace{10mm}	\textrm{Test each discretized value of } \hat{d}_k \textrm{ in range }  -r \textrm{ to } r \\
& \hspace{10mm} \beta_k = \underset{\hat{d}_k}{\text{argmin}} \|\boldsymbol{y_{L_k}} - \boldsymbol{\hat{F}_{L_k}}(\hat{d}_k)\boldsymbol{\hat{x}} \|_2 \\
&\hspace{10mm} \textbf{for each } i \textrm{ in } L_k \textbf{ do}\\
& \hspace{20mm} \textrm{Compute } \delta_i \textrm{ from } \beta_k \textrm{ using } \delta_i = h(\beta_k,u_i)
\end{align*}
In the above steps, $\boldsymbol{y_{L_k}}$ is a subvector of $\boldsymbol{y}$, containing measurements for frequencies at indices only in $L_k$, and $\boldsymbol{\hat{F}_{L_k}}(\hat{d}_k)$ denotes a sub-matrix of $\boldsymbol{\hat{F}}$ containing only those rows with indices in $L_k$ and assuming perturbation parameter $\hat{d}_k$. Note that the modification to the main algorithm essentially computes only each \emph{unique} value in $\boldsymbol{\beta}$ separately. Convergence results for Algorithm \ref{A1} (or its modification) are analyzed in Section \ref{sec:convergence}.

\subsection{Comparison with Algorithms for Basis Mismatch or DOA estimation}
\label{sec:comp_basismismatch}
We emphasize that our computational problem is \emph{very different} from the basis mismatch problem \cite{Nichols2014,Chi2011,Teke2013}. There, the signal is to be represented as a linear combination of (possibly sinsuoidal) bases whose frequencies are assumed to lie on a discrete grid, i.e. $\boldsymbol{x} = \boldsymbol{\Phi \Psi \theta} = \boldsymbol{\Phi} \sum_{k=0}^{K-1} \boldsymbol{\Psi_k} \theta_k$, where $\boldsymbol{\Psi_k} \in \mathbb{C}^N$ is the basis vector at discrete frequency $k$, and $\boldsymbol{\theta} \in \mathbb{C}^N$. However in many applications, the signals may be sparse linear combinations of bases whose frequencies lie off the grid. Hence the representation problem involves solving for the frequency perturbations $\delta_k$ along with $\boldsymbol{\theta}$ given $\boldsymbol{x}$, where $\boldsymbol{x} = \boldsymbol{\Psi_{\delta} \theta} = \sum_{k=0}^{K-1} \boldsymbol{\Psi_{\delta_k}} \theta_k$. Here $\boldsymbol{\Psi_{\delta}}$ is a perturbed form of $\boldsymbol{\Psi}$, and $\delta_k$ denotes the difference between the $k^{\textrm{th}}$ off-grid frequency and its nearest grid-point. The problem can be extended to a compressive setting, where we have measurements of the form $\boldsymbol{y} = \boldsymbol{\Phi} \sum_{k=0}^{K-1} \boldsymbol{\Psi_{\delta_k}} \theta_k$. In this (compressive) basis mismatch problem, the perturbations are in $\boldsymbol{\Psi}$ and not in $\boldsymbol{\Phi}$, unlike in our paper where the perturbations are in $\boldsymbol{\Phi}$. This leads to the following major points of difference:
\begin{enumerate}
\item In the basis mismatch problem, the number of $\delta$ values is equal to the signal dimension $N$ (or in some variants, equal to $\|\boldsymbol{\theta}\|_0$), unlike the problem in this paper where it is equal to $M$ (or $P$ if we count perturbation parameters in $\boldsymbol{\beta}$). 

\item Moreover, unless $\boldsymbol{\Phi \Psi}$ is orthonormal (which is not possible in a compressive setting), the different $\delta$ values cannot be solved through independent searches in the basis mismatch problem and require block coordinate descent for optimization. This is in contrast to the problem in this paper (See Algorithm \ref{A1} and its modification). 

\item In the basis mismatch problem, the performance is affected by the minimal separation between the components of $\boldsymbol{\theta}$ \cite{Tang2013} (and increased frequency resolution can make the problem more under-determined and increase the coherence of the matrix $\boldsymbol{\Phi \Psi}$), unlike in our problem.

\item A Taylor approximation approach in the basis mismatch problem would yield a system of equations of the form 
\begin{equation}
\boldsymbol{y} = (\boldsymbol{F} + \boldsymbol{F' \Delta}) \boldsymbol{x} + \boldsymbol{\eta_{Taylor}},
\label{eq:taylor_basismismatch}
\end{equation}
where $\boldsymbol{x}$ and $\boldsymbol{\Delta x}$ are vectors with the same support, $\boldsymbol{F}$ represents the Fourier measurement matrix at known frequency set $\{u_i\}_{i=1}^M$, $\boldsymbol{F'}$ is the first derivative of the Fourier matrix w.r.t. $\boldsymbol{\delta}$, $\boldsymbol{\Delta} \triangleq \textrm{diag}(\boldsymbol{\delta})$ and $\boldsymbol{\eta_{Taylor}}$ represents error due to truncation of the Taylor series. This allows for simultaneous estimation of $\boldsymbol{x}$ and $\boldsymbol{\Delta x}$ using joint sparsity. For our problem, the Taylor expansion leads to equations of the form:
\begin{equation}
    \boldsymbol{y} = \boldsymbol{F_{t}x} \approx (\boldsymbol{F} + \boldsymbol{\Delta}\boldsymbol{F'})\boldsymbol{x} + \boldsymbol{\eta_{Taylor}}.
 \label{eq:our_taylor}
\end{equation}
Here, we notice that even if $\boldsymbol{x}$ is sparse, the vector $\boldsymbol{F'x}$ (and hence $\boldsymbol{\Delta F'x}$) is not sparse. Hence a joint-sparsity model cannot be directly used for our problem.
\end{enumerate}

The DOA estimation techniques in \cite{zhang2012robustly,nehorai2014structured} and the synthetic aperture radar (SAR) target location estimation technique in \cite{Fannjiang2013} (see Eqns. (3) and (13) of \cite{Fannjiang2013}) are also related to the basis mismatch problem, and use the aforementioned joint sparsity. The DOA estimation technique follows the model $\boldsymbol{y} = \boldsymbol{A_{(d + \delta)} \theta}$ where $\boldsymbol{d}$ is a vector that contains parameters that represent the $N$ different grid-aligned directions. The $j^{\textrm{th}}$ column of $\boldsymbol{A_{(d + \delta)}}$ is given as $a_l(d_j, \delta_j) = \frac{1}{\sqrt{n}} \exp (\iota \pi (d_j + \delta_j) (l-(M+1)/2))$ where $l = 0,...,M-1$ and $j=0,...,N-1$ and $\iota \triangleq \sqrt{-1}$ (see for example, Section III-F of \cite{zhang2012robustly}). Here again, the number of $\delta$ values is equal to $N$ similar to the basis mismatch problem. 

\section{Empirical Results}
\label{sec:results}
\subsection{Recovery of 1-D signals}
We present recovery results on signals in a multitude of cases below, using the modified version of Algorithm \ref{A1} (i.e. with a replacement of step 9 as described in the previous section). In each chart (see Figures \ref{fig:fig1},\ref{fig:fig2},\ref{fig:fig3},\ref{fig:fig4}), 1D signals of $N = 101$ elements were used, the sparsity $s \triangleq \|\boldsymbol{x}\|_0$ of the signal was varied  along the x-axis, and the number of measurements $M$ was varied along the y-axis. The cell at the intersection depicts the relative recovery error (RRMSE), $\frac{\|\boldsymbol{x}-\boldsymbol{\hat{x}}\|_2}{\|\boldsymbol{x}\|_2}$, averaged across 5 different signals. For any sparsity level, the signals were generated using randomly chosen supports with random values at each index in the support. Thus, different signals had different supports. The base frequencies $\boldsymbol{u}$ for the $M$ Fourier compressive measurements for each signal were chosen uniformly randomly from $\{-N/2,-N/2+1,...,N/2\}$. Each base frequency was subjected to perturbations chosen from $\textrm{Uniform}[-r,+r]$, for two separate cases with $r = 1$ and $r = 0.5$ respectively. (See Section \ref{sec:probdef} for the meaning of $r$.) Note that the same $M$ base frequencies $\boldsymbol{u}$ for the Fourier sensing matrix were chosen for each signal, but the perturbations $\boldsymbol{\delta}$ were chosen differently for each signal. 
In  Figures \ref{fig:fig1},\ref{fig:fig2},\ref{fig:fig3},\ref{fig:fig4}, black (RGB (0,0,0)) indicates perfect recovery, and white (RGB (1,1,1)) indicates recovery error of $100\%$  or higher. Note that all the figures show error values plotted on the same scale, and hence the shades are comparable within and across figures. In all experiments, a multi-start strategy with 10 starts was adopted. In principle, we can avoid ambiguity in the estimation of the $\boldsymbol{\delta}$ values only if $r$ is less than half the smallest difference between the selected base frequencies. However even relaxation of this condition did not have any major adverse effect on the signal reconstruction. Note that the regularization parameter $\lambda$ in Eqn. \ref{eq:gen_prob} was chosen by cross-validation on a small `training set' of signals. The same $\lambda$ was used in all experiments. For our implementation, we used the CVX package\footnote{\url{http://cvxr.com/cvx/}}.

Figure \ref{fig:fig1} shows results for two different cases (top and bottom figures, for both $r = 1$ and $r = 0.5$): where the number of unique values in $\boldsymbol{\delta}$ are 2 and 10 respectively (this is henceforth denoted as $\delta_{(u)}$), although there are $M$ measurements. (In this experiment, the perturbation parameters in $\boldsymbol{\beta}$ are the same as the perturbation values in $\boldsymbol{\delta}$.) In both cases, no external noise was added to the measurements. One can see that the average recovery error decreases with the number of measurements and increases with $s$, although the relationship is not strictly monotonic. Figure \ref{fig:fig2} shows the same two cases as in Figure \ref{fig:fig1}, but with an addition of zero mean i.i.d. Gaussian noise with $\sigma = 5\%$ of the average magnitude of the individual (noiseless) measurements. The same trend of decrease in error with increased number of measurements and increase in error with increased $s$ is observed here as well. For reference, we also include a typical sample reconstruction in 1D canonical basis for a signal of length $101$, which is $10$-sparse, in Figure \ref{fig:1drecon}. Figure \ref{fig:fig3} shows similar results as in Figure \ref{fig:fig2} but using signals that are sparse in the Haar wavelet basis instead of the canonical basis. 

\subsection{Baselines for Recovery of 1-D signals}
\begin{figure}[!htb]
    \centering
    \subfloat[]{\includegraphics[width=4.5cm]{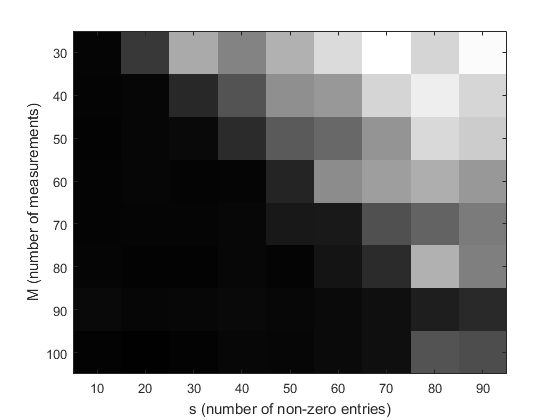}}
    \subfloat[]{\includegraphics[width=4.5cm]{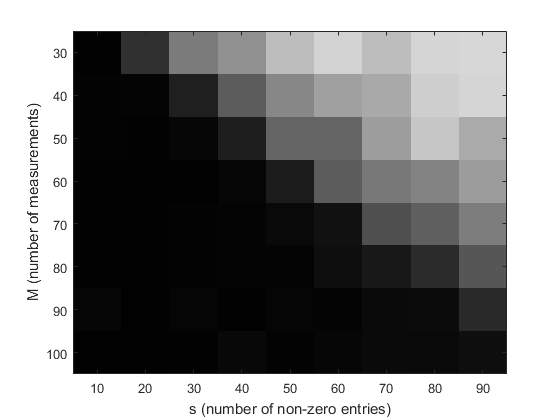}}\
      \subfloat[]{\includegraphics[width=4.5cm]{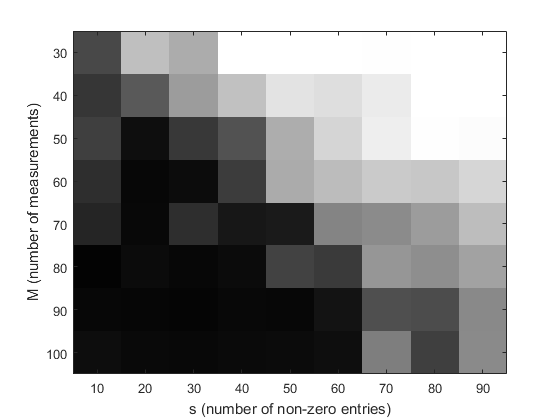}}
        \subfloat[]{\includegraphics[width=4.5cm]{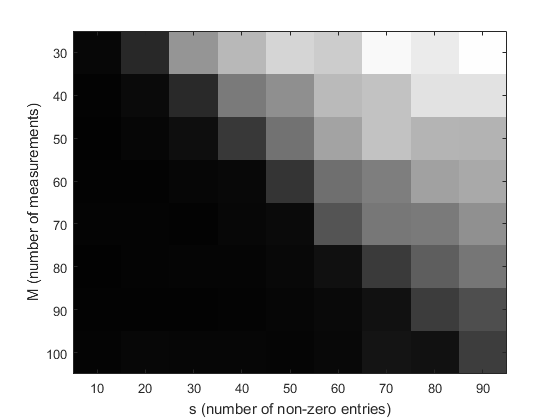}}
        \caption{Recovery with Proposed Alternating Minimization algorithm for a 1D signal with 101 elements, sparse in canonical basis, no measurement noise added (a) $r = 1$, $\delta_{(u)} = 2$, (b) $r = 0.5$, $\delta_{(u)} = 2$, (c) $r = 1$, $\delta_{(u)} = 10$, (d) $r = 0.5, \delta_{(u)} = 10$, where $\delta_{(u)}$ represents number of unique values in $\boldsymbol{\delta}$.} 
        \label{fig:fig1}

\subfloat[]{\includegraphics[width=4.5cm]{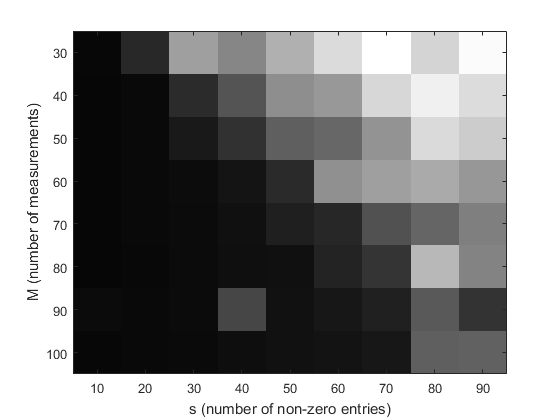}}
\subfloat[]{\includegraphics[width=4.5cm]{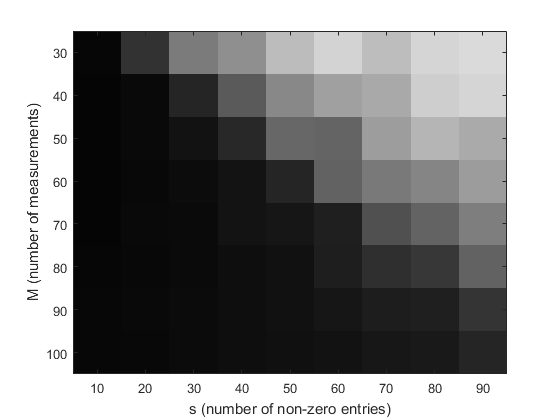}}\
\subfloat[]{\includegraphics[width=4.5cm]{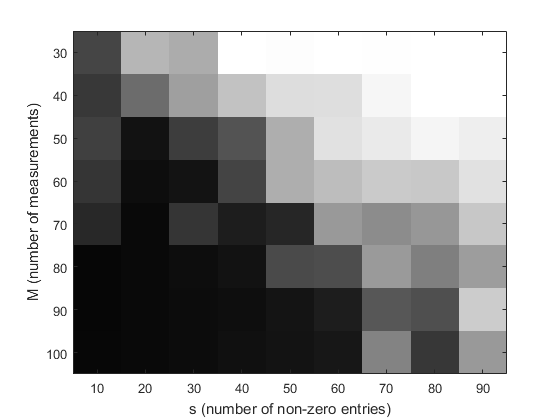}}
\subfloat[]{\includegraphics[width=4.5cm]{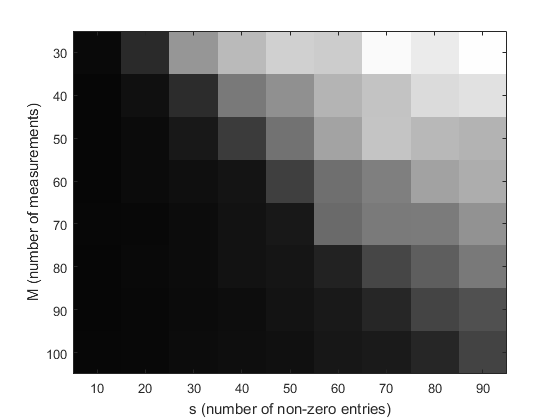}}
        \caption{Recovery with Proposed Alternating Minimization algorithm for a 1D signal with 101 elements, sparse in canonical basis, 5\% zero mean Gaussian noise added to measurements. (a) $r = 1$, $\delta_{(u)} = 2$, (b) $r = 0.5$, $\delta_{(u)} = 2$, (c) $r = 1$, $\delta_{(u)} = 10$, (d) $r = 0.5$, $\delta_{(u)} = 10$, where $\delta_{(u)}$ represents number of unique values in $\boldsymbol{\delta}$.}
        \label{fig:fig2}
\end{figure}

\begin{figure}[!htb]
    \centering
 \subfloat[]{\includegraphics[width=4.5cm]{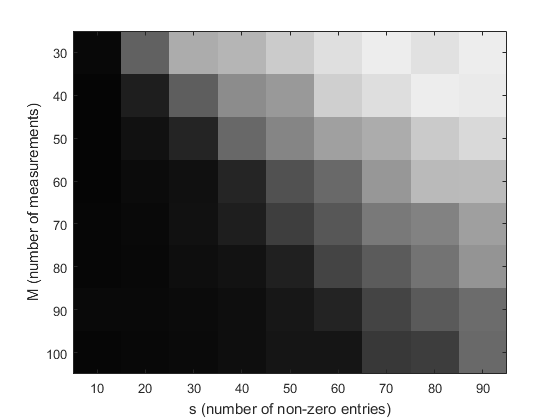}}
 \subfloat[]{\includegraphics[width=4.5cm]{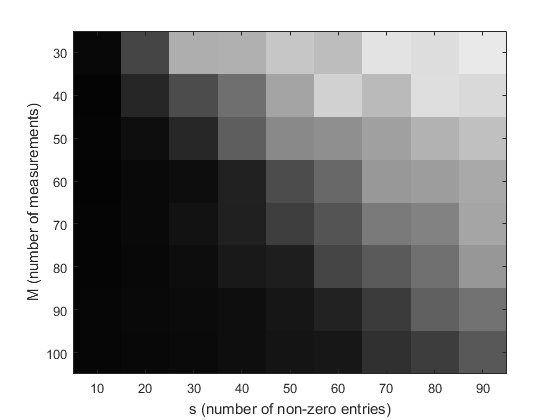}}\
 \subfloat[]{\includegraphics[width=4.5cm]{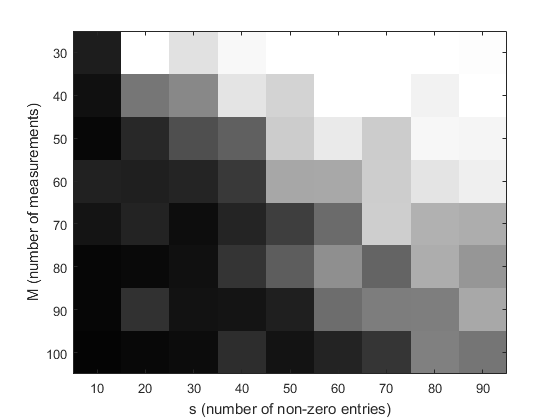}}
 \subfloat[]{\includegraphics[width=4.5cm]{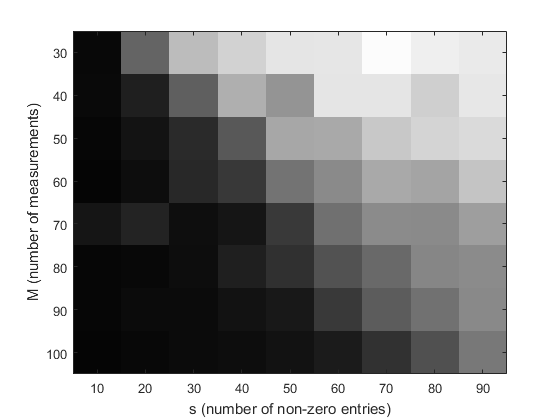}}
\caption{Recovery with Proposed Alternating Minimization algorithm for a 1D signal with 128 elements, sparse in Haar DWT basis, $5\%$ zero mean Gaussian noise added to the measurements. (a) $r = 1$, $\delta_{(u)}$ = 2 (b) $r = 0.5$, $\delta_{(u)} = 2$ (c) $r = 1$, $\delta_{(u)} = 10$, (d) $r = 0.5$, $\delta_{(u)} = 10$, where $\delta_{(u)}$ represents number of unique values in $\boldsymbol{\delta}$.}
        \label{fig:fig3}
\end{figure}

\begin{figure}[!htb]
    \centering
\subfloat[]{\includegraphics[width=4.5cm]{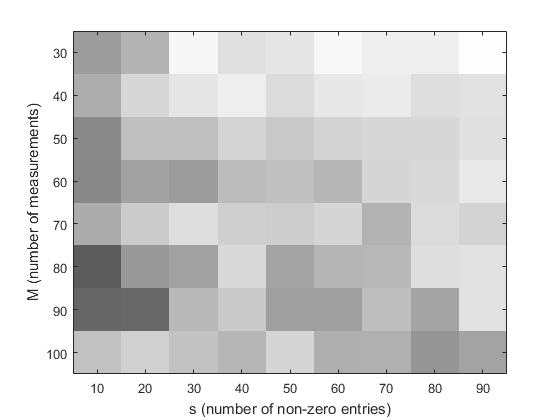}}
\subfloat[]{\includegraphics[width=4.5cm]{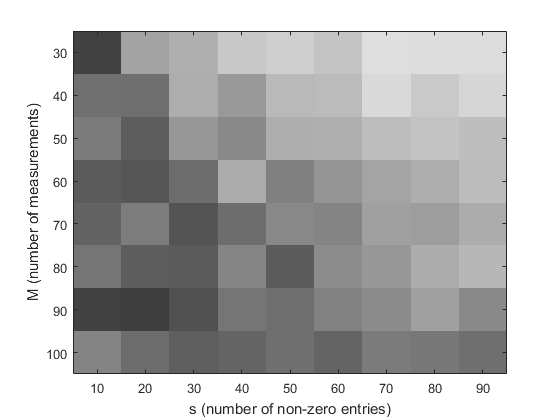}}\
\subfloat[]{\includegraphics[width=4.5cm]{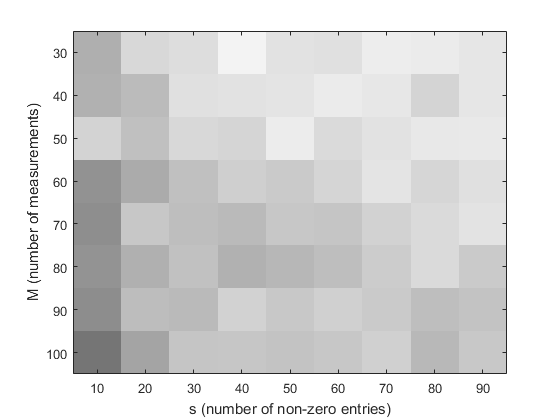}}
\subfloat[]{\includegraphics[width=4.5cm]{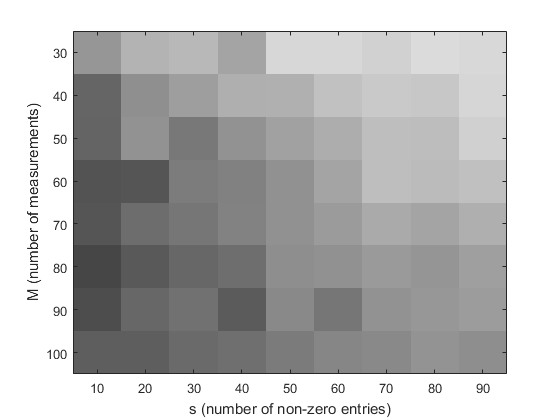}}
\caption{Recovery with Baseline 1 algorithm (see text) for a 1D signal with 101 elements, sparse in canonical basis, no measurement noise added. (a) $r = 1$, $\delta_{(u)} = 2$, (b) $r = 0.5$, $\delta_{(u)}=2$, (c) $r = 1$, $\delta_{(u)} = 10$, (d) $r = 0.5$, $\delta_{(u)}=10$, where $\delta_{(u)}$ represents number of unique values in $\boldsymbol{\delta}$. Compare to Figure \ref{fig:fig1}.}
        \label{fig:fig4}
\end{figure}

\begin{figure}[!htb]
    \centering
\subfloat[]{\includegraphics[width=4.5cm]{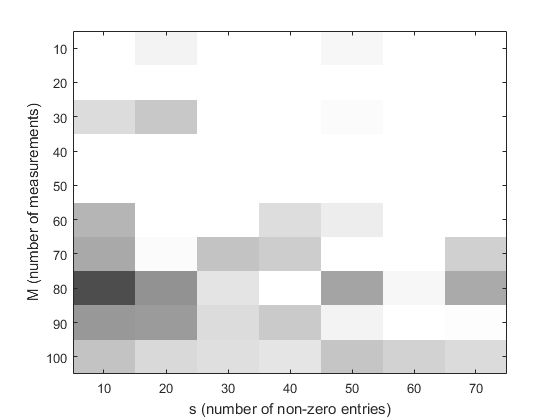}}
\subfloat[]{\includegraphics[width=4.5cm]{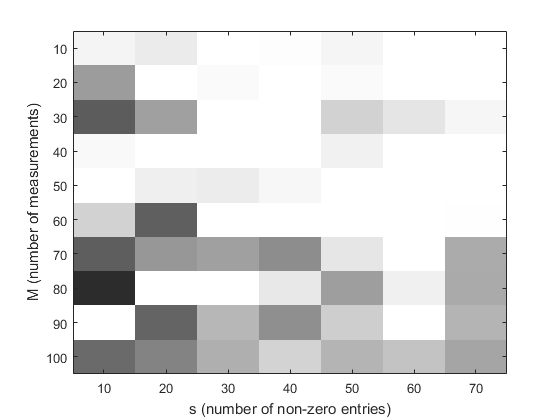}}\
\subfloat[]{\includegraphics[width=4.5cm]{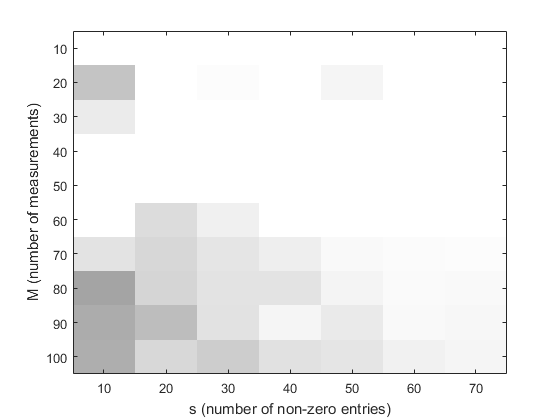}}
\subfloat[]{\includegraphics[width=4.5cm]{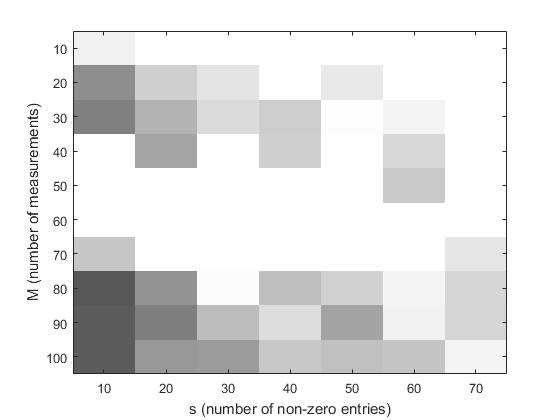}}
\caption{Recovery with Baseline 2 algorithm (see text) for a 1D signal with 101 elements, sparse in canonical basis, 5\% zero mean Gaussian noise added to measurements. (a) $r = 1$, $\delta_{(u)}=2$, (b) $r = 0.5$, $\delta_{(u)}=2$, (c) $r = 1$, $\delta_{(u)}=10$, (d) $r = 0.5$, $\delta_{(u)}=10$, where $\delta_{(u)}$ represents number of unique values in $\boldsymbol{\delta}$. Compare with Figure \ref{fig:fig2}.}
\label{fig:fig5}
\end{figure}

For comparison, we also establish two baselines: 
\begin{enumerate}
\item A naive reconstruction algorithm (termed `Baseline 1'), which ignores the perturbations and recovers the signal using a straightforward basis pursuit approach, with the \textit{unperturbed, on-grid} Fourier matrix as the measurement matrix, i.e. assuming $\boldsymbol{\delta} = \boldsymbol{0}$. Results in similar settings as in Figure \ref{fig:fig1} are shown in Figure \ref{fig:fig4}. The parameter $\lambda$ for this approach was set using cross-validation on a training set of signals.\\
\item A Taylor approximation approach (termed `Baseline 2'): Here, the signal as well as the perturbations are recovered using an alternating minimization algorithm based on a first order Taylor approximated formulation, from Eqn. \ref{eq:our_taylor}. Results in similar settings as in Figure \ref{fig:fig1} are shown in Figure \ref{fig:fig5} for two cases: one where the number of unique values in $\boldsymbol{\delta}$, , i.e. $\delta_{(u)}$, is two; and another where $\delta_{(u)} = 10$. This baseline is similar in spirit to the truncated Taylor series approach presented in \cite{nehorai2014structured,zhang2012robustly,Fannjiang2013} but modified for our (very different) computational problem. The parameter $\lambda$ for this approach was again set using cross-validation on a training set of signals.
\end{enumerate}

As is clear from the figures, Baseline 1 performs considerably worse, since inaccurate frequencies are trusted to be accurate. Baseline 2 \textit{also} performs badly because the first order Taylor error, $\boldsymbol{\eta_{Taylor}}$, can be overwhelmingly large since it is directly proportional to the unknown $\|\boldsymbol{x}\|_2$, and consequently, the signal recovered is also inferior. In fact, a comparison between Figures \ref{fig:fig4} and \ref{fig:fig5} reveals that in case of Taylor approximations to a perturbed Fourier matrix, the results obtained are often as bad as those obtained when assuming $\boldsymbol{\delta} = \boldsymbol{0}$. Baseline 2 is akin to a strategy used in \cite{nehorai2014structured,zhang2012robustly,Fannjiang2013} and applied to DOA estimation or in target detection in radar. However the specific inverse problem to be solved in these papers is similar to a problem of mismatched representation bases, which fortuitously allows for joint sparsity of $\boldsymbol{x}$ and $\boldsymbol{\delta} \cdot \boldsymbol{x}$ (see Eqn.\ref{eq:taylor_basismismatch} in Section \ref{sec:comp_basismismatch}), which cannot be achieved in the problem we attempt to solve in this paper. 



\subsection{The case of $M$ independent perturbations}
\begin{figure}[!htb]
    \centering
        \subfloat[]{\includegraphics[width=4.6cm]{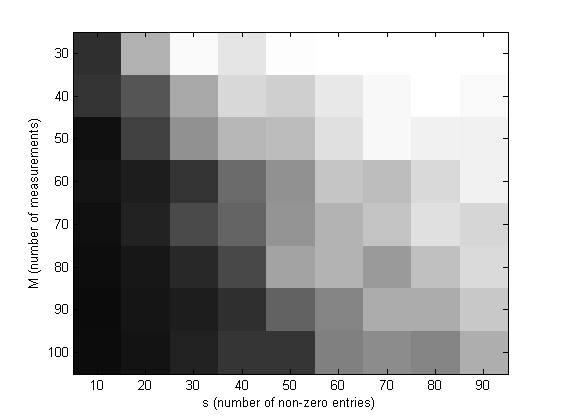}}
        \subfloat[]{\includegraphics[width=4.6cm]{results/10angles_noisy_r05.png}}
        \caption{Recovery with Proposed Alternating Minimization algorithm for a 1D signal with 128 elements, sparse in Haar DWT basis, 5\% zero mean Gaussian noise added to the measurements. Left: $r = 0.5, \delta_{(u)} = M$, where $M$ is the number of measurements, Right: $r=0.5,\delta_{(u)}=10$ - same as Fig. \ref{fig:fig2}.}
\label{fig:fig8}
\end{figure}

\begin{figure*}[!htb]
    \centering
        \includegraphics[width=\textwidth]{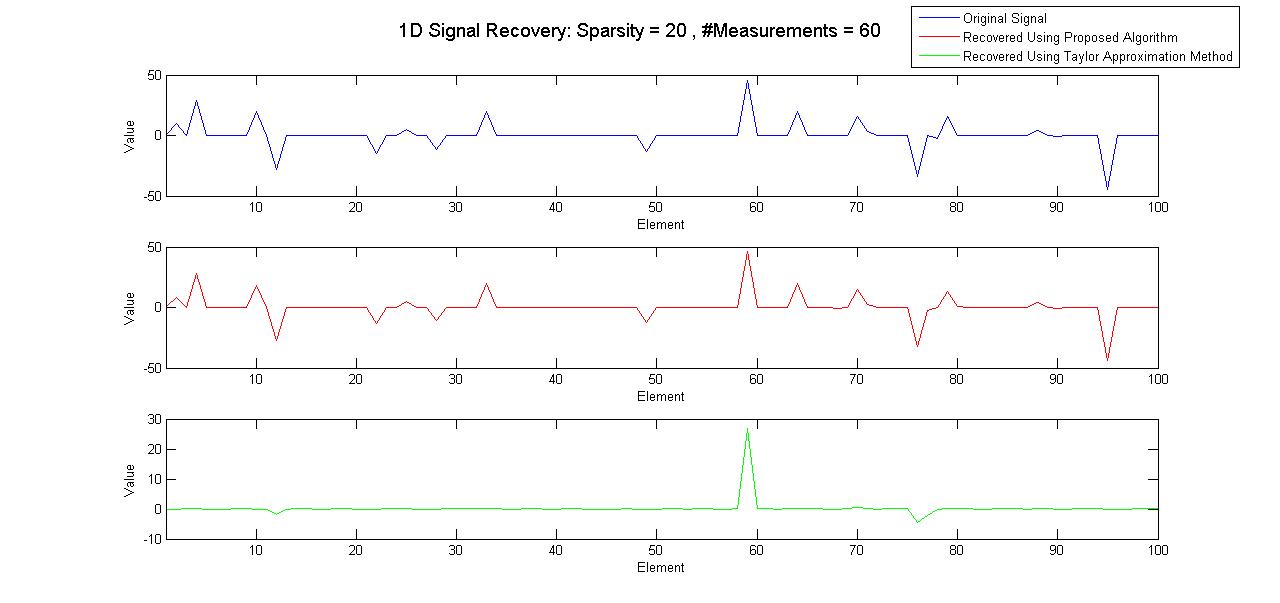}
        \caption{Sample recovery for 1D signal sparse canonical basis, $N = 100, M = 60, s = 20$, zero mean $5\%$ Gaussian noise added to measurements.  Relative reconstruction error by proposed algorithm: $5.5\%$. Relative reconstruction error by Baseline 2 (Taylor approximation): $88.7\%$.}
\label{fig:1drecon}
\end{figure*}

All the experiments so far were conducted in the setting where the number of unique values in $\boldsymbol{\delta}$ was much less than $M$. The motivation for this setting has already been described in previous sections. In the case when each measurement has an independent perturbation, we expect the recovery error to be high, especially in the presence of measurement noise, as the number of unknowns increases significantly. For completeness, we perform similar experiments in the case when $\delta_{(u)}=M$ and plot the reconstruction errors. We observe that even with large number of unique $\delta$ values, the errors are still low when the signal is very sparse (see Fig \ref{fig:fig8}). However, the error increases significantly when the signal is less sparse, and the error is much higher than the case of a small $\delta_{(u)}$ as seen in Fig. \ref{fig:fig8}.

\subsection{Recovery of 2-D signals}
\label{res}
\begin{figure}[!htb]
    \centering
        \includegraphics[width=5cm]{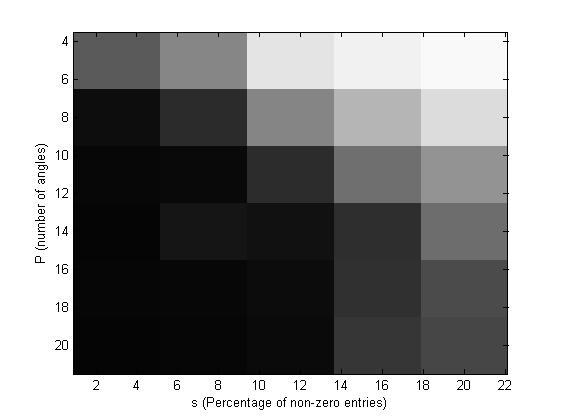}
        \caption{Recovery error for $30 \times 30$ 2D image, sparse in 2D Haar Wavelet basis, with $5\%$ zero mean Gaussian measurement noise and angle errors from $\textrm{Uniform}[-2^\circ,+2^\circ$]}
\label{fig:fig6}
\end{figure}

Application of our algorithms to 2D images is natural and more immediately applicable in imaging scenarios. We first present results with a similar set of experiments using 2D images (as the signal $\boldsymbol{x}$). For this experiment, $30 \times 30$ images were used. The images were generated using a sparse linear combination of Haar wavelet bases. We used a radial sampling approach in the Fourier domain (equivalent to taking a Fourier transform of the Radon projections), taking a fixed number of measurements along each spoke, but varying the number of angles used and the sparsity of the image in the HWT basis. The angles for the spokes were incorrectly specified (which is typical in mis-calibrated tomography) with each angle error chosen from $\textrm{Uniform}[-2^\circ,+2^\circ]$ - leading to \emph{significant} perturbations in the frequencies. The base frequencies $\boldsymbol{u}$ were spaced uniformly along each spoke. In addition, $5\%$ zero mean i.i.d. Gaussian noise was added to the measurements (both real and complex parts, independently). We used the YALL1\footnote{http://yall1.blogs.rice.edu/} solver for optimization of $\boldsymbol{x}$ and the NUFFT package\footnote{\url{https://www-user.tu-chemnitz.de/~potts/nfft/}} for computing Fourier transforms at non-integer frequencies. The results are summarized in a chart shown in Figure \ref{fig:fig6}.
As Figure \ref{fig:fig6} shows, the recovery error was small, even for a reasonably small number of measurements, and the method was robust to noise in the measurements. Errors with the baseline algorithms were significantly larger and are not reported here. 

\begin{figure}[!htb]
    \centering
        \includegraphics[width=2.7cm]{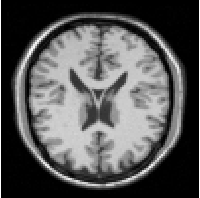}
        \includegraphics[width=2.7cm]{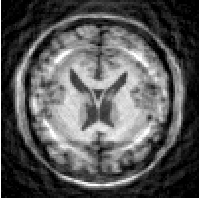}
        \includegraphics[width=2.7cm]{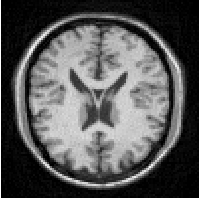}\\
        \includegraphics[width=2.7cm]{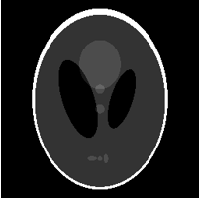}
        \includegraphics[width=2.7cm]{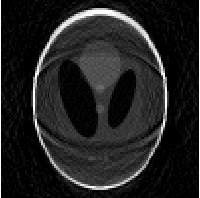}
        \includegraphics[width=2.7cm]{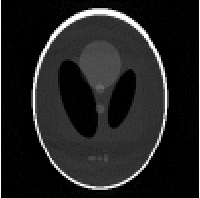}\\ \hspace{0.04em}
        \includegraphics[width=2.7cm]{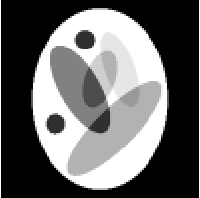}
        \includegraphics[width=2.7cm]{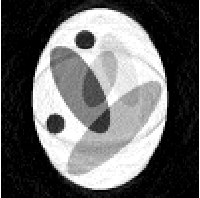} 
        \includegraphics[width=2.7cm]{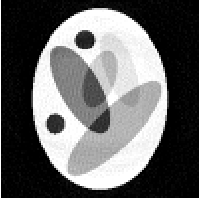}
        \caption{Reconstruction for $200 \times 200$ images with $5\%$ zero mean Gaussian measurement noise, 70\% compressive measurements, angle error from $\textrm{Uniform}[-1^\circ,+1^\circ]$. In each row, left: original image, middle: reconstruction using Baseline 1 (RRMSE $25\%,23.36\%,8.82\%$), right: reconstruction using modified version of Algorithm \ref{A1} (RRMSE $6.76\%, 5.27\%,4.5\%$).}
\label{fig:fig7}
\end{figure}

\begin{figure}[!htb]
    \centering
        \includegraphics[width=2.7cm]{reconstructions/brain_web_a.png}
        \includegraphics[width=2.7cm]{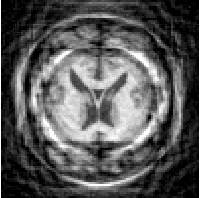}
        \includegraphics[width=2.7cm]{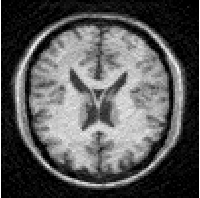}\\
        \includegraphics[width=2.7cm]{reconstructions/phantom1_a.png}
        \includegraphics[width=2.7cm]{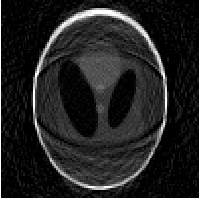}
        \includegraphics[width=2.7cm]{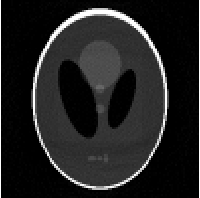}\\ \hspace{0.05em}
        \includegraphics[width=2.7cm]{reconstructions/phantom2_a.png}
        \includegraphics[width=2.7cm]{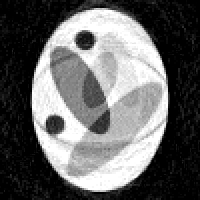} 
        \includegraphics[width=2.7cm]{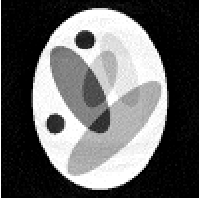}
        \caption{Reconstruction for $200 \times 200$ images with $5\%$ zero mean Gaussian measurement noise, 70\% compressive measurements, angle error from $\textrm{Uniform}[-2^\circ,+2^\circ]$. In each row, left: original image, middle: reconstruction using Baseline 1 (RRMSE $38.7\%, 30.98\%, 12.63\%$), right: reconstruction using modified version of Algorithm \ref{A1} (RRMSE $10.65\%, 5.22\%, 4.87\%$).}
\label{fig:2dimages_2degrees}
\end{figure}

\begin{figure}[!htb]
    \centering
        \includegraphics[width=2.7cm]{reconstructions/brain_web_a.png}
        \includegraphics[width=2.7cm]{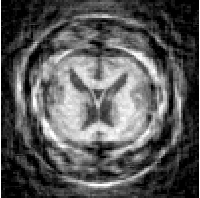}
        \includegraphics[width=2.7cm]{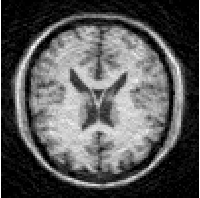}\\
        \includegraphics[width=2.7cm]{reconstructions/phantom1_a.png}
        \includegraphics[width=2.7cm]{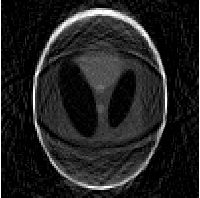}
        \includegraphics[width=2.7cm]{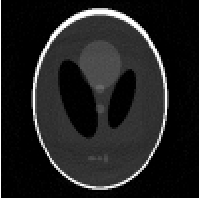}\\ \hspace{0.05em}
        \includegraphics[width=2.7cm]{reconstructions/phantom2_a.png}
        \includegraphics[width=2.7cm]{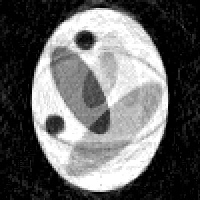}
        \includegraphics[width=2.7cm]{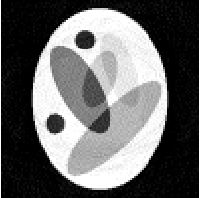}
        \caption{Reconstruction for $200 \times 200$ images with $5\%$ zero mean Gaussian measurement noise, 70\% compressive measurements, angle error from $\textrm{Uniform}[-3^\circ,+3^\circ]$. In each row, left: original image, middle: reconstruction using Baseline 1 (RRMSE $38.75\%, 35.48\%,14.59\%$), right: reconstruction using modified version of Algorithm \ref{A1} (RRMSE $13.15\%, 5.29\%, 5.85\%$).}
\label{fig:2dimages_3degrees}
\end{figure}

In the second set of experiments, we show reconstruction results on three images each of size $200 \times 200$. Fourier measurements were simulated along 140 radial spokes with erroneously specified angles (which is typical in tomography with angle errors or unknown angles). The angle error for each spoke was chosen independently from $\textrm{Uniform}[-1^\circ,+1^\circ]$ - leading to \emph{significant} perturbations in the frequencies. Noise from $\mathcal{N}(0,\sigma^2)$ where $\sigma \triangleq 0.05 \times$ average (noiseless) measurement magnitude, was added to the real and complex parts of the measurements. During reconstruction, we exploited image sparsity in a Haar wavelet basis. Reconstruction results with the modified version of Algorithm \ref{A1} are presented in Fig. \ref{fig:fig7}. In comparison with Baseline 1, we see that our algorithm performs significantly better in terms of RRMSE values as well as visually - see Fig. \ref{fig:fig7}. Results with a similar experiment for angle errors chosen independently from $\textrm{Uniform}[-2^\circ,+2^\circ]$ and $\textrm{Uniform}[-3^\circ,+3^\circ]$ are shown in Fig. \ref{fig:2dimages_2degrees} and Fig. \ref{fig:2dimages_3degrees} respectively, showing clear performance improvement of our method over Baseline 1. 

\section{Theoretical Results}
\label{sec:theory}
While the empirical results show the algorithm working well across a large number of simulated scenarios, we also characterize the formulation by providing theoretical analysis for (A) convergence of the algorithm, (B) uniqueness of the solution to the main problem, (C) the uniqueness of the minimum of a linearized approximation of the main objective function, (D) the effect of perturbations on the sensing matrix, and (E) the quality of the solution that would be obtained by ignoring perturbations. 

\subsection{Convergence of Algorithm \ref{A1}}
\label{sec:convergence}
Here we provide a proof of convergence of Algorithm \ref{A1} (or its modified version) under a specific condition mentioned further. Let $\boldsymbol{F_\delta}$ denote the Fourier transform computed at the frequencies values $\boldsymbol{u}+\boldsymbol{\delta}$ where $\boldsymbol{\delta} = h(\boldsymbol{\beta},\boldsymbol{u})$. Assign $\boldsymbol{z} = \{\boldsymbol{x}, \boldsymbol{\beta}\}$. Recall that our objective is to determine the solution $\boldsymbol{z^{*}}$ that minimizes the objective function $J(\boldsymbol{z}) \triangleq \|\boldsymbol{x}\|_1 + \lambda \|\boldsymbol{y} - \boldsymbol{F}(\boldsymbol{\delta}) \boldsymbol{x}\|_2$, namely $\boldsymbol{z^{*}} = \textrm{argmin}_{\boldsymbol{z}} J(\boldsymbol{z})$.

Let $\boldsymbol{z_t} = \{\boldsymbol{x_t}, \boldsymbol{\beta_t}\}$ be the present solution of our alternating search algorithm at iteration $t$. Our alternating search algorithm ensures that the sequence of function values $\{J(\boldsymbol{z_t})\}_{t \in \mathbb{N}}$ is monotonically decreasing. As $J$ is bounded below by $0$, the sequence $\{J(\boldsymbol{z_t})\}_{t \in \mathbb{N}}$ converges to a limit value $E \in \mathbb{R}^+$ by the monotone convergence theorem. 

\begin{figure}
\centering
\includegraphics[width=7cm]{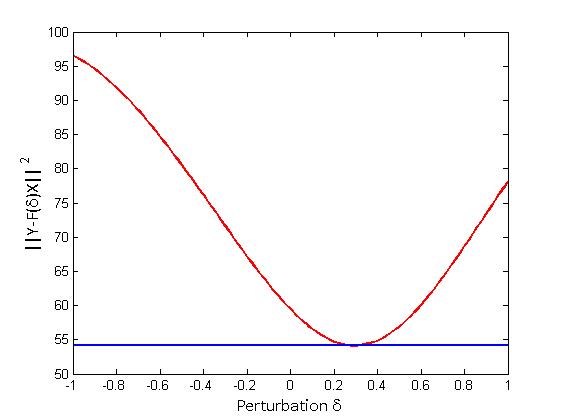}
\caption{Uniqueness of the solution for $\boldsymbol{\delta}$ keeping $\boldsymbol{x}$ fixed, where $\boldsymbol{x}$ is the estimated signal at empirically observed convergence of Algorithm \ref{A1}.}
\label{fig:energy}
\end{figure}

However, this does not yet prove the convergence of the solution sequence $\{\boldsymbol{z_t}\}$. To this end, let $\boldsymbol{x}(\boldsymbol{\beta})$ denote the minimizer for the convex objective function on $\boldsymbol{x}$ with $\boldsymbol{\beta}$ held fixed, namely $\boldsymbol{x}(\boldsymbol{\beta}) = \textrm{argmin}_{\bx} J_{\boldsymbol{\beta}}(\bx)$, where $J_{\boldsymbol{\beta}}(\bx) = J(\bz)$ with $\boldsymbol{\beta}$ held constant. In the context of our alternating search algorithm, we have $\bx_{t+1}= \bx(\boldsymbol{\beta}_t)$. Letting $\bz_{t+\frac{1}{2}} = \{\bx_{t+1}, \boldsymbol{\beta}_{t}\}$ we find
\begin{align*}
\|\bx_{t+1}\|_2 &\leq \|\bx_{t+1}\|_1 \leq  J\left(\bz_{t+\frac{1}{2}}\right) \\
						& = J_{\boldsymbol{\beta}_t}\left(\bx_{t+1}\right) \leq J_{\boldsymbol{\beta}_t}(\bzero) = \lambda \|\by\|_2
\end{align*}
giving an upper bound on the norm of $\bx_{t}$. The last but one inequality follows from that fact that $\bx_{t+1}$ minimizes $J_{\boldsymbol{\beta}_t}(\bx)$. Further, as $-r \leq \beta_i \leq r$ for each $i$, we see that the sequence $\{\bz_t\}_{t \in \mathbb{N}}$ lie within a compact space. Hence as per Theorem 4.9 in \cite{Gorski2007}, this sequence has atleast one accumulation point. Another statement in the same theorem states that if a certain condition is satisfied, then $\textrm{lim}_{t \rightarrow \infty} \|\bz_{t+1}-\bz_t\| = 0$, which establishes convergence of the solution. The condition is that for each such accumulation point, the minimization of $J(\boldsymbol{z})$ gives (i) a unique solution for $\boldsymbol{x}$ if $\boldsymbol{\beta}$ is fixed, \textrm{and} (ii) a unique solution for $\boldsymbol{\beta}$ if $\boldsymbol{x}$ is fixed. Condition (i) is easy to satisfy as the problem is convex in $\boldsymbol{x}$ if $\boldsymbol{\beta}$ is fixed. We do not have a proof for Condition (ii), but we have observed uniqueness in practice, especially since the values in $\boldsymbol{\beta}$ are bounded between $-r$ to $+r$. As an example, in Fig. \ref{fig:energy}, we show a plot of the function $\|\boldsymbol{y}-\boldsymbol{F_{\delta}x}\|^2$ keeping $\boldsymbol{x}$ and all but one value in $\boldsymbol{\delta}$ fixed. Note that here $\boldsymbol{x}$ denotes the estimated signal value upon (empirically observed) convergence of Algorithm \ref{A1}. We would like to emphasize that Theorem 4.9 in \cite{Gorski2007} only requires continuity of the function $J$ and no other conditions like biconvexity. Thus, we have established the following Lemma for conditional convergence of Algorithm \ref{A1} (and its modification) to a local minimum of $J$. Given the non-convexity of $J$, global guarantees are very difficult to establish. 
\begin{lemma}
\label{lemma:conv}
Algorithm \ref{A1} is locally convergent if for every accumulation point of the sequence $\boldsymbol{z_t}$, Condition (ii) is satisfied.
\end{lemma}

\subsection{Uniqueness of Solution}
\label{sec:uniqueness}
It is quite natural to question whether the recovery of $\boldsymbol{x}$ from compressive measurements of the form $\boldsymbol{y} = \boldsymbol{F_t x}$ is unique, where $\boldsymbol{F_t}$ is as defined in Eqn. \ref{eq:eq1}. We answer this question in the affirmative (in the noiseless case, of course) under the condition that the perturbation parameters $\boldsymbol{\beta}$ be independent of the base frequencies $\boldsymbol{u}$, i.e. $\forall i, 1 \leq i \leq P, \delta_i = \tilde{h}(\beta_i)$ where $\tilde{h}$ is a known function of only $\beta_i$. We comment on the effect of relaxing this condition, at the end of the section. \\
First consider real-valued $\boldsymbol{x}$, which is typical in tomography and certain protocols in MR (if the magnetization is proportional to the contrast-weighted proton density \cite{kspace_wiki}). Consider the case where there is only a single unknown perturbation parameter value $\beta$ in all measurements and where $\boldsymbol{x} $ is a 1D signal. Then, we have:
\begin{equation}
\boldsymbol{y} = \boldsymbol{F_t x} = \boldsymbol{F} (\boldsymbol{x} \cdot \boldsymbol{v}_{\beta}),
\end{equation}
where $\boldsymbol{v}_{\beta}$ is a vector in $\mathbb{C}^N$ whose $l^{\textrm{th}}$ entry is equal to $\exp(-\iota 2\pi \tilde{h}(\beta) l / N)$ where $l$ is a spatial/time index and $\iota = \sqrt{-1}$. To see this, consider the $i^{\textrm{th}}$ measurement as follows:
\begin{eqnarray}
y_i = \frac{1}{\sqrt{M}} \sum_l \exp(-\iota 2 \pi (u + \delta) l /N) x(l) \\ \nonumber
= \frac{1}{\sqrt{M}} \sum_l \exp(-\iota 2 \pi u l /N) (x(l) \exp(-\iota 2 \pi \tilde{h}(\beta) l /N)).
\end{eqnarray}
Let $\boldsymbol{x}_{\beta} \triangleq \boldsymbol{x} \cdot \boldsymbol{v}_{\beta}$. Using standard compressive sensing results from \cite{Candes2008}, we can prove unique recovery of $\boldsymbol{x}_{\beta}$ using basis pursuit, for sufficiently large $M$ ($M \geq s \log N$ for $s$-sparse $\boldsymbol{x}$) and an RIP-obeying $\boldsymbol{F}$ (which is true if the base frequencies were chosen uniformly at random \cite{Rudelson2008}). Since $\boldsymbol{x}$ is real-valued, both $\boldsymbol{x}$ and $\beta$ are uniquely recovered. Moreover this recovery is robust to measurement noise and compressibility (instead of strict sparsity) of $\boldsymbol{x_{\beta}}$ and the bounds from \cite{Candes2008} would follow. This result extends to $\boldsymbol{x}$ in higher dimensions as well. If $\boldsymbol{x} \in \mathbb{C}^N$, although $\boldsymbol{x}_{\beta}$ can be recovered uniquely, there is an inevitable phase ambiguity in estimating $\boldsymbol{x}$. By the Fourier shift theorem, this implies that $\boldsymbol{x}$ can be estimated only up to a global shift, which depends upon $\beta$. However, the magnitude of each element of $\boldsymbol{x}$, i.e. $|\boldsymbol{x}|$, can be recovered uniquely under the afore-stated conditions.

Consider the case of $P \ll M$ unique perturbation parameter values in $\boldsymbol{\beta}$, $\boldsymbol{x} \in \mathbb{R}^N$, and $\boldsymbol{x}$ is $s$-sparse. Let $\boldsymbol{F_{L_k}}$ be the sub-matrix of measurements corresponding to a particular value $\beta_k$. Using the earlier arguments, unique recovery of $\boldsymbol{x}, \boldsymbol{\delta}$ can be guaranteed if for at least one $k \in \{1,2,...,P\}$, the matrix $\boldsymbol{F_{L_k}}$ obeys the RIP of order $s$. If $\boldsymbol{x} \in \mathbb{C}^N$, then one can guarantee unique recovery of only $|\boldsymbol{x}|$.

These uniqueness results can be further strengthened (i.e. in terms of weaker conditions on the number of measurements), by observing that in the case of $P > 1$ unique values in $\boldsymbol{\beta}$, we need to recover different (complex) signals $\boldsymbol{x}_{\beta_1},\boldsymbol{x}_{\beta_2},...,\boldsymbol{x}_{\beta_P}$ where $\forall i, 1 \leq i \leq P, \boldsymbol{x}_{\beta_i} \triangleq \boldsymbol{x} \cdot \boldsymbol{v}_{\beta_i}$ and the $l^{th}$ entry of $\boldsymbol{v}_{\beta_i}$ equals $\exp(-\iota 2\pi \tilde{h}(\beta_i) l/N)$. All these signals are $s$-sparse if $\boldsymbol{x}$ is $s$-sparse, and they have the same support. This recovery problem is therefore an example of multiple measurement vectors (MMV), for which stronger recovery results exist - see Theorem 18 of \cite{Duarte2011}. However, our computational problem has further refinements to MMV: the sensing sub-matrices corresponding to the different values in $\boldsymbol{\beta}$ are necessarily different, which is termed the generalized MMV (GMMV) problem \cite{Heckel2012},\cite{Rajamohan2017}, for which stronger results exist. For example, we modify Theorem 1 of \cite{Heckel2012} which guarantees unique recovery of the sparsity pattern of the signals, to state the following Lemma:
\begin{lemma}
Consider measurements $\forall k, 1 \leq k \leq P, \boldsymbol{y_{L_k}} = \boldsymbol{F_{L_k}} \boldsymbol{x}_{\beta_k}$ where $\boldsymbol{x}_{\beta_k} = \boldsymbol{x} \cdot \boldsymbol{v}_{\beta_k}$ and $v_{\beta_k}(l) = \exp(-\iota 2 \pi \tilde{h}(\beta_k) l/N)$. Assume that $\boldsymbol{x}$ is $s$-sparse with support set denoted $\mathcal{S}$ and has sub-Gaussian entries. Assume that the following conditions hold:
\begin{eqnarray}
\forall j \notin \mathcal{S}, \Big(\frac{1}{P} \sum_{k=1}^P \|\boldsymbol{F^{\dagger}_{L_k,\mathcal{S}}} \boldsymbol{F_{L_k,j}}\|^2 \Big)^{0.5} \leq \alpha_1 < 1 \\
\forall j \notin \mathcal{S}, \textrm{max}_{k \in \{1,...,P\}} \|\boldsymbol{F^{\dagger}_{L_k,\mathcal{S}}} \boldsymbol{F_{L_k,j}}\|_2 \leq \alpha_2 > 0,
\end{eqnarray}
where $\dagger$ denotes the pseudo-inverse, $\boldsymbol{F_{L_k,j}}$ is the $j^{\textrm{th}}$ column of $\boldsymbol{F_{L_k}}$ and $\boldsymbol{F_{L_k,\mathcal{S}}}$ is a sub-matrix of $\boldsymbol{F_{L_k}}$ with columns corresponding to entries in $\mathcal{S}$. Then the solution to the following optimization problem (Q1) is able to recover the exact solution for the signals $\boldsymbol{x}_{\beta_1},\boldsymbol{x}_{\beta_2},...,\boldsymbol{x}_{\beta_P}$ with high probability decreasing in $\alpha_1, \alpha_2$. The problem (Q1) is defined as follows:
$\textrm{min} \|\boldsymbol{x}\|_1 \textrm{ s. t. } \forall k \in \{1,...,P\} \textrm{ } \boldsymbol{y_{L_k}} = \boldsymbol{F_{L_k}} \boldsymbol{x}_{\beta_k}$. \QEDA
\end{lemma}
Clearly, GMMV results require weaker conditions than MMV (see eqn. 12 of \cite{Heckel2012}). However, our computational problem in fact has further structure over and above GMMV. First, $\forall i, 1 \leq i \leq N, |\boldsymbol{x}_{\beta_1}(i)| = |\boldsymbol{x}_{\beta_2}(i)| =...= |\boldsymbol{x}_{\beta_P}(i)|$. Second, the phase factors of all elements of $\boldsymbol{x}_{\beta_1},\boldsymbol{x}_{\beta_2},...,\boldsymbol{x}_{\beta_P}$ are \emph{completely determined} by just the $P$ values in $\boldsymbol{\beta}$. The modified version of Algorithm \ref{A1} imposes this structure by design. At this point, we conjecture that the lower bound on the required number of measurements is actually much lower, if we use Algorithm \ref{A1} for estimation of $\boldsymbol{x}, \boldsymbol{\beta}$, as compared to the predictions from the aforementioned CS, MMV, GMMV approaches. Moreover, we conjecture that Algorithm \ref{A1} is also more robust to measurement noise by design, as compared to these approaches. 

Lastly, we consider the case when the values in $\boldsymbol{\delta}$ are functions of the base frequencies in addition to the values in $\boldsymbol{\beta}$ (i.e. $\forall i \in \{1,2,...,M\}, \exists! k \in \{1,2,...,P\} \textrm{ s. t. } \delta_i = h(\beta_k,u_i)$ where $\exists!$ is the unique existential quantifier), which is more challenging. This is because it requires estimation of $M$ (as opposed to $P$) signals, albeit all with common support and with the aforementioned structure. Empirically however, we have observed success of Algorithm \ref{A1} even in such a scenario (see Fig. \ref{fig:fig7}).

\subsection{Theoretical Analysis for a Linearized Approximation}
The analysis in the previous section requires that at least one measurement sub-matrix $\boldsymbol{F_{L_k}}$ (corresponding to a given perturbation parameter $\beta_k$) obeys the RIP. The analysis does not hold in the case where $P = M$. As such, theoretical error bounds for the global optimum of Algorithm \ref{A1} are difficult due to the fact that the perturbations $\hat{\bdelta}$ feature non-linearly inside the Fourier matrix $\hat{\boldsymbol{F}}$. Therefore, we set out to analyze a linearized measurement model given in the statement of Theorem \ref{thm:taylor} below, which is applicable in the $P = M$ case. Even for such an approximation and with $M = N$, the analysis is far from simple and the uniqueness of a solution is not obvious. The sole purpose of Theorem \ref{thm:taylor} is to establish that there exists a unique solution in the linearized noiseless setting. We hope that Theorem \ref{thm:taylor} will pave the way for future research for obtaining uniqueness results in the general non-linear case.

In the following, we consider $\boldsymbol{\Delta} \triangleq \textrm{diag}(\boldsymbol{\delta})$. Let $\boldsymbol{F}$ denote the Fourier matrix at frequencies $\boldsymbol{u}$. We can treat $\boldsymbol{F_{t}}$ as approximated by $\boldsymbol{F} + \boldsymbol{\Delta} \boldsymbol{F'}$, where $\boldsymbol{F'} = \boldsymbol{FX}$ is the derivative of the Fourier matrix with respect to the elements in $\boldsymbol{\Delta}$ and $\boldsymbol{X}$ is a diagonal matrix, with $\boldsymbol{X}_{ll} = \frac{2 \pi l}{N}$ where $l$ is the index to the spatial location ranging from $-(N+1)/2 \leq l \leq (N+1)/2$. In other words, $\boldsymbol{F_{t}} \approx \boldsymbol{F} + \boldsymbol{\Delta} \boldsymbol{F'}$. Without loss of generality we assume $N$ is odd. We now state and prove the following theorem:
\begin{theo}
\label{thm:taylor}
For measurements $\boldsymbol{y} \in \mathbb{C}^M$ of the form 
\begin{equation*}
    \boldsymbol{y} = (\boldsymbol{F} + \boldsymbol{\Delta }\boldsymbol{F}\boldsymbol{X})\boldsymbol{x},
    \label{eq:taylor_thm}
\end{equation*}
the signal $\boldsymbol{x} \in \mathbb{R}^N$ and the perturbations $\boldsymbol{\Delta}$ (diagonal $N \times N$ matrix) can both be uniquely recovered with probability $1$, independent of the sparsity of the signal $\boldsymbol{x}$ and the magnitude of the values in $\boldsymbol{\Delta}$, if (a) $M = N$,  (b) $\boldsymbol{x}$ is neither purely even nor purely odd, (c) $\boldsymbol{y}$ does not contain any pair of elements that are conjugate symmetric, and (d) the frequencies in $\boldsymbol{u}$ form an anti-symmetric set such that $u_{(M+1)/2 - k} = - u_{(M+1)/2 + k}$ for $1 \leq k \leq \frac{M-1}{2}$. \QEDA
\end{theo}
\textit{Proof:} To prove this, we perform a series of non-trivial algebraic manipulations to arrive at a linear system of the form $\boldsymbol{g} = \boldsymbol{Hw}$ where $\boldsymbol{w}$ is related purely to $\boldsymbol{x}$, and $\boldsymbol{g},\boldsymbol{H}$ depend only upon $\boldsymbol{y}, \boldsymbol{F}$. The uniqueness of the solution then follows by showing the invertibility of $\boldsymbol{H}$ (with high probability) in the $M = N$ case. We comment upon the $M < N$ case thereafter. During the proof for the $M=N$ case, we also precisely point out why the four assumptions (a)-(d) in the Theorem statement are required, and argue that they are weak assumptions. 

We utilize the following defined relations:
\begin{equation}
\boldsymbol{F} \triangleq \boldsymbol{C} + \iota \boldsymbol{S} , \boldsymbol{x} \triangleq \boldsymbol{e} + \boldsymbol{o}, \boldsymbol{y} \triangleq \boldsymbol{y_r} + \iota \boldsymbol{y_c}
\end{equation}
where, $\boldsymbol{C}$ is the cosine component and $\boldsymbol{S}$ is the sine component of $\boldsymbol{F}$, $\boldsymbol{e}$ is the even component of $\boldsymbol{x}$ (having $N$ elements), and $\boldsymbol{o}$ is the odd component of $\boldsymbol{x}$ (having $N$ elements), $\boldsymbol{y_r}$ is the real component of $\boldsymbol{y}$, and $\iota \boldsymbol{y_c}$ is the complex component of $\boldsymbol{y}$. Given $\boldsymbol{x}$, we note that $\boldsymbol{e}$ and $\boldsymbol{o}$ are uniquely defined.

Therefore, $\boldsymbol{y} = (\boldsymbol{F} + \boldsymbol{\Delta }\boldsymbol{F}\boldsymbol{X})\boldsymbol{x}$ can be rewritten as:
\begin{equation}
\boldsymbol{y_r} + \iota\boldsymbol{y_c} = \boldsymbol{C}\boldsymbol{x} + \iota\boldsymbol{S}\boldsymbol{x} + \iota\boldsymbol{\Delta}(\boldsymbol{C} + \iota\boldsymbol{S})\boldsymbol{X}\boldsymbol{x},
\end{equation}
which implies that
\begin{align}
\boldsymbol{y_r} &= \boldsymbol{C x} -\boldsymbol{\Delta S X x}, \mbox{  and}
\label{eq:yryc1} \\
\boldsymbol{y_c}& = \boldsymbol{S}\boldsymbol{x} + \boldsymbol{\Delta} \boldsymbol{C}\boldsymbol{X}\boldsymbol{x}.
\label{eq:yryc2}
\end{align}
Note that $\boldsymbol{Co} = \boldsymbol{S}\boldsymbol{Xo} = \boldsymbol{Se} = \boldsymbol{C}\boldsymbol{Xe} = \boldsymbol{0}$. \\
We now divide $\boldsymbol{C}$ into smaller submatrices: The central column, $\boldsymbol{C}_{0}$, and the left \& right components, $\boldsymbol{C}_{-1}$ and $\boldsymbol{C}_{1}$ respectively.
That is, \\
let
$\boldsymbol{C} = \begin{bmatrix} \boldsymbol{C}_{-1}       & \boldsymbol{C}_{0} & \boldsymbol{C}_{1} \end{bmatrix} =  \begin{bmatrix}
    \boldsymbol{C}_{-1,-1}       & \boldsymbol{C}_{0,-1} & \boldsymbol{C}_{1,-1} \\
	\boldsymbol{C}_{-1,0}       & \boldsymbol{C}_{0,0} & \boldsymbol{C}_{1,0} \\    
    \boldsymbol{C}_{-1,1}       & \boldsymbol{C}_{0,1} & \boldsymbol{C}_{1,1}
\end{bmatrix}$. \\ 
where the second division is done in a similar fashion, along the row axis. To summarise, 
$\boldsymbol{C}_{0,0}$ is a single centre element, $\boldsymbol{C}_{-1,0}$ and $\boldsymbol{C}_{1,0}$ are row vectors, $\boldsymbol{C}_{0,-1}$ and $\boldsymbol{C}_{0,1}$ are column vectors, and finally $\boldsymbol{C}_{-1,-1}$, $\boldsymbol{C}_{1,-1}$, $\boldsymbol{C}_{-1,1}$ and $\boldsymbol{C}_{1,1}$ are matrices of size $(M-1)/2 \times (N-1)/2$.
\\ Similarly, let \\
$\boldsymbol{S} = \begin{bmatrix} \boldsymbol{S}_{-1}       & \boldsymbol{S}_{0} & \boldsymbol{S}_{1} \end{bmatrix} =  \begin{bmatrix}
    \boldsymbol{S}_{-1,-1}       & \boldsymbol{S}_{0,-1} & \boldsymbol{S}_{1,-1} \\
	\boldsymbol{S}_{-1,0}       & \boldsymbol{S}_{0,0} & \boldsymbol{S}_{1,0} \\    
    \boldsymbol{S}_{-1,1}       & \boldsymbol{S}_{0,1} & \boldsymbol{S}_{1,1},
\end{bmatrix}$
where $\boldsymbol{S_{-1}},\boldsymbol{S_1}, \boldsymbol{S_0}$ are similarly defined. 

Consider the case that the frequencies in $\boldsymbol{u}$ form an anti-symmetric set about $0$, which is stated in assumption (d). Note that this is true for on-grid frequencies from $-(N+1)/2$ to $(N+1)/2$ used in a typical DFT matrix, or in applications such as radial/Cartesian MRI or CT. Then, we further have
\begin{equation*}
\begin{matrix}
\boldsymbol{C}_{-1} = \boldsymbol{C}_{1}, &
\boldsymbol{S}_{-1} = -\boldsymbol{S}_{1}, \\
\boldsymbol{C}_{-1,-1} = \boldsymbol{C}_{-1,1}, &
\boldsymbol{S}_{-1,-1} = -\boldsymbol{S}_{-1,1}, \\
\boldsymbol{C}_{1,-1} = \boldsymbol{C}_{1,1}, &
\boldsymbol{S}_{1,-1} = -\boldsymbol{S}_{1,1},
\end{matrix}
\end{equation*}
\begin{equation*}
\begin{matrix}
\boldsymbol{C}_{0} = \begin{bmatrix}
1 \\ 1 \\ \vdots \\ 1
\end{bmatrix}, \mbox{  and}&
\boldsymbol{S}_{0} = \begin{bmatrix}
0 \\ 0 \\  \vdots \\ 0
\end{bmatrix}.
\end{matrix}
\end{equation*}
Separating out $\boldsymbol{e}$ and $\boldsymbol{o}$  and $\boldsymbol{X}$ also in a similar fashion, let $\boldsymbol{e} = \begin{bmatrix}
    \boldsymbol{e}_{-1} \\
	\boldsymbol{e}_{0}  \\    
    \boldsymbol{e}_{1}
\end{bmatrix}$, $\boldsymbol{o} = \begin{bmatrix}
    \boldsymbol{o}_{-1} \\
	\boldsymbol{o}_{0}  \\    
    \boldsymbol{o}_{1}
\end{bmatrix}$
and $\boldsymbol{X} = \begin{bmatrix}
    \boldsymbol{X}_{-1} & 0 &  \boldsymbol{0}\\
	 \boldsymbol{0} & 0 &  \boldsymbol{0} \\
     \boldsymbol{0} & 0 &  \boldsymbol{X}_{1}
\end{bmatrix}$
where $\boldsymbol{e}_{-1}=\boldsymbol{e}_{1}$, $ \boldsymbol{o}_{-1} = - \boldsymbol{o}_{1}$ and $\boldsymbol{X}_{-1} = - \boldsymbol{X}_{1}$. 
Using these, equations~\ref{eq:yryc1} and \ref{eq:yryc2} can be rewritten as: 
\begin{align}
\label{eq:yrimproved}
\boldsymbol{y_r} &= 2\boldsymbol{C}_1\boldsymbol{e}_1 + \boldsymbol{C}_0\boldsymbol{e}_0 - \boldsymbol{\Delta}(2\boldsymbol{S}_1\boldsymbol{X}_1\boldsymbol{e}_1) \\
\boldsymbol{y_c} &= 2\boldsymbol{S}_1\boldsymbol{o}_1 + \boldsymbol{\Delta}(2\boldsymbol{C}_1\boldsymbol{X}_1\boldsymbol{o}_1).
\end{align}
When $\boldsymbol{x}$ is purely even ($\boldsymbol{o_1=0}$) or purely odd ($\boldsymbol{e_1=0}$), then $\boldsymbol{y_c}$ or $\boldsymbol{y_r}$ is respectively zero and we have $N+\lceil N/2 \rceil$ unknown quantities to solve using $N$ known values of either $\boldsymbol{y_r}$ or $\boldsymbol{y_c}$. Clearly, we do not have uniqueness. Consequently, we require assumption (b) in the theorem. (Note also that this assumption is a weak one, as most signals encountered in practice are neither purely even nor purely odd.) Using~\ref{eq:yrimproved}, the middle component ${y}_{r0}$ is given by:
\begin{equation*}
\boldsymbol{y}_{r0} = 2\boldsymbol{C}_{1,0}\boldsymbol{e}_1 + \boldsymbol{C}_{0,0}\boldsymbol{e}_0 - \boldsymbol{\Delta}_0(2\boldsymbol{S}_{1,0}\boldsymbol{X}_1\boldsymbol{e}_1).
\end{equation*}
As $\boldsymbol{S}_{1,0} = 0$ we get $
\boldsymbol{C}_{0,0}\boldsymbol{e}_0 = \boldsymbol{y}_{r0} - 2\boldsymbol{C}_{1,0}\boldsymbol{e}_1$.
Substituting this in equation~\ref{eq:yrimproved}, we get
\begin{equation*}
\boldsymbol{y_r} - \boldsymbol{1} \boldsymbol{y}_{r0} = 2(\boldsymbol{C}_1 - \boldsymbol{\Delta} \boldsymbol{C}_{1,0})\boldsymbol{e}_1 - 2\boldsymbol{\Delta} \boldsymbol{S}_1\boldsymbol{X}_1\boldsymbol{e}_1
\end{equation*}
where $\boldsymbol{1}$ is a column vector of $1$s.
Define the quantities
\begin{equation*}
\begin{matrix}
\boldsymbol{C}_r \triangleq 2(\boldsymbol{C}_1 - \boldsymbol{\Delta} \boldsymbol{C}_{1,0}), &
\boldsymbol{S}_r \triangleq \boldsymbol{S}_c = 2\boldsymbol{S}_1, & \\
\boldsymbol{C}_c \triangleq 2\boldsymbol{C}_1, &
\boldsymbol{a} \triangleq \boldsymbol{y_r} - \boldsymbol{1} \boldsymbol{y}_{r0}, &
\boldsymbol{b} \triangleq \boldsymbol{y_c}.
\end{matrix}
\end{equation*}
Further, write $\boldsymbol{a} \triangleq \begin{bmatrix}
    \boldsymbol{a}_{-1} \\
	0  \\    
    \boldsymbol{a}_{1}
\end{bmatrix}$, $\boldsymbol{b} \triangleq \begin{bmatrix}
    \boldsymbol{b}_{-1} \\
	\boldsymbol{b}_{0}  \\    
    -\boldsymbol{b}_{1}
\end{bmatrix}$ and \\
$\boldsymbol{\Delta} \triangleq \begin{bmatrix}
    \boldsymbol{\Delta}_{-1} & 0 &  \boldsymbol{0}\\
	 \boldsymbol{0} & \boldsymbol{\Delta}_{0} &  \boldsymbol{0} \\
     \boldsymbol{0} & 0 &  \boldsymbol{\Delta}_{1}
\end{bmatrix}$.
We then obtain the reduced set of equations:
\begin{align}
\label{key1} \boldsymbol{a}_{-1} &= \boldsymbol{C}_{r,-1}\boldsymbol{e}_1 - \boldsymbol{\Delta}_{-1}\boldsymbol{S}_{r,-1}\boldsymbol{X}_1\boldsymbol{e}_1 \\
\label{key2} \boldsymbol{a}_{1} &= \boldsymbol{C}_{r,1}\boldsymbol{e}_1 - \boldsymbol{\Delta}_{1}\boldsymbol{S}_{r,1}\boldsymbol{X}_1\boldsymbol{e}_1 \\
\label{key3} \boldsymbol{b}_{-1} &= \boldsymbol{S}_{c,-1}\boldsymbol{o}_1 + \boldsymbol{\Delta}_{-1}\boldsymbol{C}_{c,-1}\boldsymbol{X}_1\boldsymbol{o}_1 \\
\label{key4} -\boldsymbol{b}_{1} &= \boldsymbol{S}_{c,1}\boldsymbol{o}_1 + \boldsymbol{\Delta}_{1}\boldsymbol{C}_{c,1}\boldsymbol{X}_1\boldsymbol{o}_1 
\end{align}
where $\boldsymbol{C}_{r,-1}=\boldsymbol{C}_{r,1}$, $\boldsymbol{S}_{r,-1}=-\boldsymbol{S}_{r,1}$, $\boldsymbol{C}_{c,-1}=\boldsymbol{C}_{c,1}$ and $\boldsymbol{S}_{c,-1}=-\boldsymbol{S}_{c,1}$. Ergo,
\begin{align*}
\boldsymbol{a}_{1} - \boldsymbol{a}_{-1} & = -(\boldsymbol{\Delta}_{1} + \boldsymbol{\Delta}_{-1})\boldsymbol{S}_{r,1}\boldsymbol{X}_1\boldsymbol{e}_1 \\
\boldsymbol{b}_{1} - \boldsymbol{b}_{-1} & = -(\boldsymbol{\Delta}_{1} + \boldsymbol{\Delta}_{-1})\boldsymbol{C}_{c,1}\boldsymbol{X}_1\boldsymbol{o}_1.
\end{align*}
Let $\boldsymbol{\Sigma} \triangleq -(\boldsymbol{\Delta}_{1} + \boldsymbol{\Delta}_{-1})$
\begin{align}
\implies  
\boldsymbol{S}_{r,1}\boldsymbol{X}_1\boldsymbol{e}_1 &= \boldsymbol{\Sigma}^{-1}(\boldsymbol{a}_{1} - \boldsymbol{a}_{-1}) \\
\boldsymbol{C}_{c,1}\boldsymbol{X}_1\boldsymbol{o}_1 &= \boldsymbol{\Sigma}^{-1}(\boldsymbol{b}_{1} - \boldsymbol{b}_{-1}).
\end{align}
It can be verified that the assumption (c) in the theorem is equivalent to presuming that $\boldsymbol{\Delta}_{-1,k} \neq -\boldsymbol{\Delta}_{1,k}$ for any diagonal component $k$ and thus $\boldsymbol{\Sigma}$ is invertible. This assumption is again a weak one, as in most applications such as CT, radial/Cartesian MRI, the perturbations will usually obey $\boldsymbol{\Delta}_{-1,k} \neq -\boldsymbol{\Delta}_{1,k}$. Since $\boldsymbol{\Sigma}$ is diagonal we have
\begin{align}
\boldsymbol{S}_{r,1}\boldsymbol{X}_1\boldsymbol{e}_1 &= \textrm{diag}(\boldsymbol{a}_{1} - \boldsymbol{a}_{-1})\textrm{diag}(\boldsymbol{b}_{1} - \boldsymbol{b}_{-1})^{-1}\boldsymbol{\Sigma}^{-1}(\boldsymbol{b}_{1} - \boldsymbol{b}_{-1}) \nonumber \\
 &= \textrm{diag}(\boldsymbol{a}_{1} - \boldsymbol{a}_{-1})\textrm{diag}(\boldsymbol{b}_{1} - \boldsymbol{b}_{-1})^{-1} \boldsymbol{C}_{c,1}\boldsymbol{X}_1\boldsymbol{o}_1 \nonumber \\
 \label{eq:M1}
 &=\boldsymbol{Z} \boldsymbol{C}_{c,1}\boldsymbol{X}_1\boldsymbol{o}_1 
\end{align}
where $\boldsymbol{Z} = \textrm{diag}(\boldsymbol{a}_{1} - \boldsymbol{a}_{-1})\textrm{diag}(\boldsymbol{b}_{1} - \boldsymbol{b}_{-1})^{-1}$. As both $\boldsymbol{Z}$ and $\boldsymbol{\Delta}_{-1}$ are diagonal matrices, they commute and we get 
\begin{equation*}
\boldsymbol{\Delta}_{-1}\boldsymbol{S}_{r,1}\boldsymbol{X}_1\boldsymbol{e}_1 = \boldsymbol{Z} \boldsymbol{\Delta}_{-1} \boldsymbol{C}_{c,1}\boldsymbol{X}_1\boldsymbol{o}_1 = \boldsymbol{Z} \left[\boldsymbol{b}_{-1} +\boldsymbol{S}_{c,1}\boldsymbol{o}_1 \right]
\end{equation*}
where the last equality follows from equation~\ref{key3} and the relation $\boldsymbol{S}_{c,-1}=-\boldsymbol{S}_{c,1}$.
Substituting in equation~\ref{key1} gives us:
\begin{eqnarray}
\boldsymbol{a}_{-1} = \boldsymbol{C}_{r,1}\boldsymbol{e}_1 + \boldsymbol{Z}\boldsymbol{b}_{-1} + \boldsymbol{Z}\boldsymbol{S}_{c,1}\boldsymbol{o}_1
\label{eq:M2} \\
\implies \boldsymbol{a}_{-1} - \boldsymbol{Z}\boldsymbol{b}_{-1}  = \boldsymbol{C}_{r,1}\boldsymbol{e}_1 + \boldsymbol{Z}\boldsymbol{S}_{c,1}\boldsymbol{o}_1.
\end{eqnarray}
Consider equations \ref{eq:M1} and \ref{eq:M2}. These can be written in matrix form as:
\begin{equation}
\label{eq:ghw}
\begin{bmatrix}
\boldsymbol{a}_{-1} - \boldsymbol{Z}\boldsymbol{b}_{-1} \\ \boldsymbol{0}
\end{bmatrix}
=
\begin{bmatrix}
\boldsymbol{C}_{r,1} & \boldsymbol{Z}\boldsymbol{S}_{c,1} \\
\boldsymbol{S}_{r,1}\boldsymbol{X}_1 & -\boldsymbol{Z}\boldsymbol{C}_{c,1}\boldsymbol{X}_1
\end{bmatrix}
\begin{bmatrix}
\boldsymbol{e}_1 \\ \boldsymbol{o}_1
\end{bmatrix}.
\end{equation}
This is of the form $ \boldsymbol{g} = \boldsymbol{H} \boldsymbol{w}$, with 
\begin{equation*}
\boldsymbol{g} \triangleq \begin{bmatrix}
\boldsymbol{a}_{-1} - \boldsymbol{Z}\boldsymbol{b}_{-1} \\ \boldsymbol{0}
\end{bmatrix},
\boldsymbol{H} \triangleq \begin{bmatrix}
\boldsymbol{C}_{r,1} & \boldsymbol{Z}\boldsymbol{S}_{c,1} \\
\boldsymbol{S}_{r,1}\boldsymbol{X}_1 & -\boldsymbol{Z}\boldsymbol{C}_{c,1}\boldsymbol{X}_1
\end{bmatrix},
\boldsymbol{w} \triangleq \begin{bmatrix}
\boldsymbol{e}_1 \\ \boldsymbol{o}_1
\end{bmatrix}.
\end{equation*}
Using this linear system of equations, we can recover $\boldsymbol{w}$. With $M = N$, this is a simple case of inverting the $\boldsymbol{H}$ matrix. 
Since $\boldsymbol{w} = \begin{bmatrix}
\boldsymbol{e}_1 \\ \boldsymbol{o}_1
\end{bmatrix}$, recovering $\boldsymbol{w}$ immediately gives us the signal $\boldsymbol{x}$. If $M = N$ (assumption (a)), the applicability of the previous results hinges on the invertibility of $\boldsymbol{H}$. We can show that $\boldsymbol{e_1}$ and $\boldsymbol{o_1}$ are indeed necessarily recoverable from this linear system.

Consider the second set of equations in \ref{eq:ghw}. i.e. $\boldsymbol{S}_{r,1} \boldsymbol{X}_1 \boldsymbol{e}_1 = \boldsymbol{Z}\boldsymbol{C}_{c,1} \boldsymbol{X}_1 \boldsymbol{o}_1$, which implies that $\boldsymbol{e}_1 = \boldsymbol{X}_1^{-1}\boldsymbol{S}_{r,1}^{-1}\boldsymbol{Z}\boldsymbol{C}_{c,1}\boldsymbol{X}_1\boldsymbol{o}_1$. All the matrices involved here are composed of elementary entries and are invertible because of the properties of the Fourier matrix. Substituting this value of $\boldsymbol{e}_1$ in the first set of equations in \ref{eq:ghw}, we get
\begin{eqnarray}
\boldsymbol{a}_{-1} - \boldsymbol{Z}\boldsymbol{b}_{-1} = [\boldsymbol{C}_{r,1} \boldsymbol{X}_1^{-1} \boldsymbol{S}_{r,1}^{-1} \boldsymbol{Z} \boldsymbol{C}_{c,1} \boldsymbol{X}_1 + \boldsymbol{Z} \boldsymbol{S}_{c,1} ] \boldsymbol{o}_1 \\
\implies \boldsymbol{S}_{c,1}^{-1}\boldsymbol{Z}^{-1} (\boldsymbol{a}_{-1} - \boldsymbol{Z} \boldsymbol{b}_{-1}) = (\boldsymbol{E} + \boldsymbol{I}) \boldsymbol{o}_1,
\end{eqnarray}
where 
\begin{equation*}
\boldsymbol{E} \triangleq \boldsymbol{S}_{c,1}^{-1}\boldsymbol{Z}^{-1} \boldsymbol{C}_{r,1} \boldsymbol{X}_1^{-1}\boldsymbol{S}_{r,1}^{-1} \boldsymbol{Z} \boldsymbol{C}_{c,1} \boldsymbol{X}_1.
\end{equation*}  
Let $\boldsymbol{A} \triangleq \boldsymbol{I},  \boldsymbol{U} \triangleq \boldsymbol{S}_{c,1}^{-1}\boldsymbol{Z}^{-1},  \boldsymbol{C} \triangleq \boldsymbol{C}_{r,1} \boldsymbol{X}_1^{-1} \boldsymbol{S}_{r,1}^{-1}, \boldsymbol{V} \triangleq \boldsymbol{Z} \boldsymbol{C}_{c,1} \boldsymbol{X}_1$. Then $\boldsymbol{E} + \boldsymbol{I} = \boldsymbol{A} + \boldsymbol{U C V}$. The measurement $\boldsymbol{Z}$ is independent of both $\boldsymbol{A}$ and $\boldsymbol{C}$. For the matrix $\boldsymbol{E} + \boldsymbol{I}$ to not be invertible, we would need to select a precise $\boldsymbol{Z}$, so as to get an eigenvalue of $-1$ for $\boldsymbol{E}$. Given that the perturbations are picked uniformly at random, the matrix is invertible with probability $1$. This shows that using the approximation as described, the signal $\boldsymbol{x}$ is recoverable uniquely with high probability, when $M = N$. \QEDA

\begin{figure}[!htb]
    \centering
        \includegraphics[width=8cm]{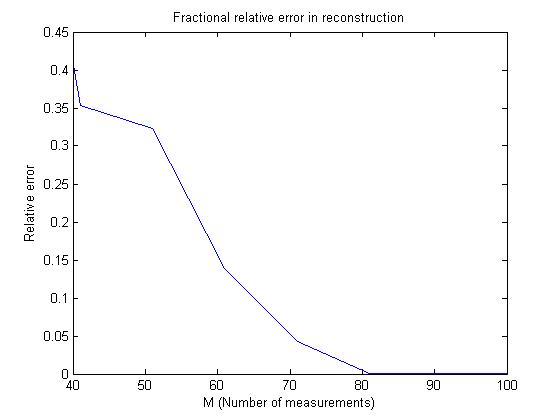}
        \caption{Relative recovery error using linearized model formulation}
        \label{fig:linearmodelerr}
\end{figure}

While we haven't shown a bound in a  compressed sensing framework, where $M < N$, we have empirically observed that recovery is excellent in this scenario as well. Empirically, we observe good recovery with this formulation, even with a row-subsampled Fourier matrix ($M \leq N$). The variation of the error is presented for a 20-sparse signal of length 100 in figure \ref{fig:linearmodelerr}. The reported error is with respect to measurements simulated by the linearized approximation model $\boldsymbol{y} = (\boldsymbol{F} + \boldsymbol{\Delta }\boldsymbol{F}\boldsymbol{X})\boldsymbol{x}$ as per Theorem \ref{thm:taylor}.

\subsection{Coherence of Perturbed Fourier Matrix}
\label{subsec:coh_pfm}
There exist results from the compressed sensing literature that derive performance bounds on signal reconstruction in terms of the mutual coherence of the sensing matrix $\boldsymbol{\Phi}$ \cite{Studer2014}. We provide a bound on the mutual coherence (hereafter simply referred to as `coherence') in the expected sense assuming that the perturbations are random draws from $\textrm{Uniform}[-r,+r]$, and subsequently prove this bound. \\

\begin{theo}
\label{thm:coherence}
Let $\boldsymbol{F_t}$ be the Fourier matrix at frequencies $\boldsymbol{u} + \boldsymbol{\delta}$, where $\boldsymbol{u}$ represents (the possibly but not necessarily on-grid) frequency set, and $\boldsymbol{\delta} \sim \textrm{Uniform}[-r,+r], r > 0$ represents the perturbation to this set. Let the unperturbed Fourier matrix at frequencies $\boldsymbol{u}$ be denoted by $\boldsymbol{F}$. Then, the expected coherence $\mu_t \triangleq \left\|E(\boldsymbol{\Psi}^T \boldsymbol{F_t}^\mathsf{H} \boldsymbol{F_t} \boldsymbol{\Psi})\right\|_{\infty}$ respects the inequality
\begin{equation*}
\mu_t \leq \underset{\theta}{\text{max}} \mu \frac{|\sin(\theta r)|}{\theta r} \leq \mu
\end{equation*}
where $\boldsymbol{\Psi}$ is a fixed orthonormal signal-representation matrix, $\mu$ is the coherence of $\boldsymbol{F\Psi}$, and $\theta$ takes values $\frac{2\pi (j_1 - j_2)}{N}$, for $j_1 \neq j_2, j_1 \in \{0,1,...,N-1\}, j_2 \in \{0,1,...,N-1\}$. \QEDA
\end{theo}
\textit{Proof:} To bound the expected coherence, we will use the assumption that each $\delta_k$ is drawn i.i.d. from $\textrm{Uniform}[-r,+r]$. Consider the matrix $\boldsymbol{V} = \boldsymbol{\Psi}^\mathsf{T} \boldsymbol{F_t}^\mathsf{H} \boldsymbol{F_t} \boldsymbol{\Psi}$ whose entires equal $V_{j_1,j_2} = \boldsymbol{\Psi^t_{j_1}} (\boldsymbol{F_t})^\mathsf{H} \boldsymbol{F_t} \boldsymbol{\Psi_{j_2}}$ where $\boldsymbol{\Psi_{j_1}}$ represents the column of $\boldsymbol{\Psi}$ at index $j_1$. Then we have the following:
\begin{equation}
E[V_{j_1,j_2}] = \boldsymbol{\Psi^t_{j_1}} E[(\boldsymbol{F_t})^\mathsf{H} \boldsymbol{F_t}] \boldsymbol{\Psi_{j_2}}. 
\label{eq:EVj1j2}
\end{equation}
We define $\boldsymbol{B} \triangleq E[(\boldsymbol{F_t})^\mathsf{H} \boldsymbol{F_t}]$ and let  $\theta_{j_1,j_2} \triangleq 2 \pi (j_1-j_2)/N$. Using the uniform distribution of each $\delta_k$ we find
\begin{align*}
B_{j_1,j_2} &= \int_{-r}^{r} \frac{1}{M} \sum_k  \exp\left( \iota \theta_{j_1,j_2} (u_k+\delta_k)\right) \frac{1}{2r} d\delta_k \\
 & = \frac{1}{M} \sum_k  \frac{\exp\left( \iota \theta_{j_1,j_2} u_k\right)}{2r} \int_{-r}^{r} \exp\left( \iota \theta_{j_1,j_2} \delta_k \right) d\delta_k.
\end{align*}
Furthermore, since we know that
\begin{equation*}
\int_{-r}^{r} \sin(\theta_{j_1,j_2} \delta_k) d\delta_k = 0, 
\end{equation*}
and
\begin{equation*}
\int_{-r}^{r} \cos(\theta_{j_1,j_2} \delta_k) d\delta_k = \frac{2}{\theta_{j_1,j_2}}\sin(\theta_{j_1,j_2} r),
\end{equation*}
we get
\begin{align}
B_{j_1,j_2} &= \frac{1}{N} \sum_k  \frac{\exp\left( \iota \theta_{j_1,j_2} u_k\right)}{2r} \frac{2}{\theta_{j_1,j_2}}\sin(\theta_{j_1,j_2} r) \\
&= \dfrac{\sin (\theta_{j_1,j_2} r)}{\theta_{j_1,j_2} r} \left( \dfrac{1}{N} \sum_{k} \exp (\iota \theta_{j_1,j_2} u_k) \right). 
\label{eq:EV}
\end{align}

Let $\boldsymbol{C}$ be a matrix such that $C_{j_1,j_2} \triangleq \frac{1}{N} \sum_{k} \exp (\iota \theta_{j_1,j_2} u_k)$.
Then 
\begin{equation}
B_{j_1,j_2} = \dfrac{\sin (\theta_{j_1,j_2} r)}{\theta_{j_1,j_2} r} \left( C_{j_1,j_2} \right). 
\end{equation}
Substituting back in equation \ref{eq:EVj1j2},
\begin{align}
E[V_{j_1,j_2}] &= |\boldsymbol{\Psi^t_{j_1}} \boldsymbol{B} \boldsymbol{\Psi_{j_2}}| \\
 &= |\boldsymbol{\Psi^t_{j_1}} \dfrac{\sin (\theta_{j_1,j_2} r)}{\theta_{j_1,j_2} r} \boldsymbol{C} \boldsymbol{\Psi_{j_2}}| \\
 &= |\left( \dfrac{\sin (\theta_{j_1,j_2} r)}{\theta_{j_1,j_2} r} \right) \boldsymbol{\Psi^t_{j_1}} \boldsymbol{C} \boldsymbol{\Psi_{j_2}}| .  
\end{align}

By the definition of $\boldsymbol{C}$, we see that $\textrm{max}_{j_1,j_2,j_1 \neq j_2} |\boldsymbol{\Psi^t_{j_1} C \Psi_{j_2}} | \leq \mu$. This yields us the following:
\begin{equation}
E[V_{j_1,j_2}] = |\boldsymbol{\Psi^t_{j_1} B \Psi_{j_2}} | \leq \Big|\frac{\sin(\theta_{j_1,j_2} r)}{r\theta_{j_1,j_2}}\Big| \Big|\boldsymbol{\Psi^t_{j_1} C \Psi_{j_2}}\Big|.
\end{equation}
Since the last quantity on the RHS is nothing but the coherence $\mu$, this further yields,
\begin{equation}
E(V_{j_1,j_2}) \leq \mu \Big|\frac{\sin(\theta_{j_1,j_2} r)}{r\theta_{j_1,j_2}}\Big|.
\end{equation}
Since $\Big|\frac{\sin(\theta_{j_1,j_2} r)}{\theta_{j_1,j_2} r}\Big|$ is the absolute value of a sinc function, it takes a maximum value of $1$. Therefore, the expected coherence is less than or equal to the coherence of $\boldsymbol{F \Psi}$. In practice, the coherence values of the two matrices were found to be extremely close. \QEDA
\\
If $\boldsymbol{\Psi} = \boldsymbol{I}$, then $\boldsymbol{F \Psi}$ has low coherence with high probability if the frequencies are chosen uniformly at random \cite{Candes2006}. Hence, in this case, we can call upon well-established compressive sensing results \cite{Studer2014,Candes2006} to show that our problem is well-founded in theory. 

\subsection{Bound on Recovery in Expectation}
Assuming the frequency perturbations are obtained i.i.d. from $\textrm{Uniform}[-r,+r]$, consider the expected measurement vector from the system $\boldsymbol{F_t} \boldsymbol{x} + \boldsymbol{\eta}$ where $\boldsymbol{\eta}$ represents a measurement noise vector. That is,
\begin{equation}
\tilde{\boldsymbol{y}} = \mathbb{E}_\Delta[\boldsymbol{y}] + \boldsymbol{\eta}
= \mathbb{E}_\Delta[\boldsymbol{F_t}\boldsymbol{x}] + \boldsymbol{\eta}.
\end{equation}
We now show a bound on the error in reconstructing $\boldsymbol{x}$ from $\boldsymbol{\tilde{y}}$ if the unperturbed Fourier matrix $\boldsymbol{F}$ were to be used during reconstruction, i.e. if one simply assumed the perturbations to be all equal to zero. For this, we invoke results and proof methodology from \cite{Herman2010}. 

Since each perturbation $\Delta_k$ is assumed to be independent, we can calculate the value of $\mathbb{E}_{\Delta_k}[(\boldsymbol{F_t}\boldsymbol{x})_k]$. 
\begin{align*}
&\mathbb{E}_{\Delta_k} \left[(\boldsymbol{F_t}\boldsymbol{x})_{k}\right] \\
& \hspace{10mm}	= \mathbb{E}_{\Delta_k}\left[ \frac{1}{\sqrt{M}} \sum_j \exp(\frac{-2 \pi \iota j(u_k+\Delta_k)}{N}) \boldsymbol{x}_j \right] \\
& \hspace{10mm}	= \mathbb{E}_{\Delta_k}\left[\frac{1}{\sqrt{M}} \sum_j \exp(\frac{-2 \pi  \iota j u_k}{N}) \boldsymbol{x}_j \exp(\frac{-2 \pi \iota j\Delta_k}{N})\right] \\
& \hspace{10mm}	= \left[\frac{1}{\sqrt{M}} \sum_j \exp(\frac{-2 \pi \iota j u_k}{N}) \boldsymbol{x}_j \frac{\sin \frac {2 \pi j r}{N}}{\frac{2 \pi j r}{N}} \right].
\end{align*}
The last equality follows the steps from Section \ref{subsec:coh_pfm}. Defining $\boldsymbol{G}$ to be a diagonal matrix such that $\boldsymbol{G}_{jj} \triangleq \dfrac{\sin \frac{2 \pi j r}{N}}{\frac{2 \pi j r}{N}}$, we can see that
\begin{equation}
\mathbb{E}_\Delta[\boldsymbol{F_t}\boldsymbol{x}] = \boldsymbol{F}\boldsymbol{G}\boldsymbol{x}.
\label{eq:EFtx}
\end{equation}

Using this result, we can concretely state the recovery bound in the following theorem:
\begin{theo}
\label{thm:recov}
Let $\tilde{\boldsymbol{y}} \triangleq \boldsymbol{F} \boldsymbol{G}\boldsymbol{x} + \boldsymbol{\eta}$
and $\boldsymbol{x^*}$ be the solution to the recovery problem
\begin{align*}
    &\textrm{min } & \|\boldsymbol{x}\|_1 \hspace{8mm}\\
    &\textrm{s.t. } & \|\tilde{\boldsymbol{y}}-\boldsymbol{F}\boldsymbol{x}\|_2 \leq \epsilon'
\end{align*}
where $\boldsymbol{F}$ is the Fourier matrix at unperturbed frequencies. Under mild conditions of $\bF$ and $\boldsymbol{x}$ as assumed in Theorem~2 of \cite{Herman2010}, there exists a suitable value of $\epsilon^{\prime}$ and constants $C_0$, $C_1$, for which the recovery error is bounded as:
\begin{equation*}
	\|\boldsymbol{x^*} - \boldsymbol{x}\|_2  \leq 
    \frac{C_0}{\sqrt{s}} \|\boldsymbol{x} - \boldsymbol{x_{(s)}}\|_2 + C_1 \epsilon^{\prime}
\end{equation*}
where $\boldsymbol{x_{(s)}}$ is the best $s-$term approximation of $\boldsymbol{x}$ containing the largest $s$ coefficients of $\boldsymbol{x}$ with the rest set to zero. \QEDA
\end{theo}

We now prove this result drawing upon the proof of Theorem~2 of \cite{Herman2010}, to which our formulation is analogous. To see the analogy more clearly, we define the following quantities:
\begin{align*}
\bE \triangleq \bF(\bG-\bI), \epsilon_F \triangleq \|\boldsymbol{G} - \boldsymbol{I}\|_2.
\end{align*}
We know that $\|\boldsymbol{F}\|_2$ equals the largest singular value of matrix $\bF$. Let $\|\boldsymbol{F}\|_2^{(s)}$ denote the largest singular value taken
over all $s$-column sub-matrices of $F$. We have
\begin{equation}
\label{eqn:bound}
	\frac{\|\bE\|_2}{\|\bF\|_2} \leq \frac{\|\boldsymbol{F}\|_2 \|\boldsymbol{G}-\boldsymbol{I}\|_2}{\|\boldsymbol{F}\|_2} = \|\boldsymbol{G}-\boldsymbol{I}\|_2 = \epsilon_{\bF}.
\end{equation}
Further, since $\boldsymbol{G}$ and $\boldsymbol{I}$ are both diagonal matrices, multiplication by $(\boldsymbol{G}-\boldsymbol{I})$ conserves the sparsity of any $s$-sparse signal $\boldsymbol{x_s}$. Therefore,
$\|\bE\|_2^{(s)} \leq \|\bF\|_2^{(s)} \|\bG-\bI\|_2$ and hence $\dfrac{\|\bE\|_2^{(s)}}{\|\bF\|_2^{(s)}} \leq \epsilon_{\bF}$. 
From here, we follow exactly along the lines of the proof of Theorem~2 in \cite{Herman2010}, and arrive at a value of $\epsilon^{\prime}$ (see Eqn. 14 of \cite{Herman2010}) satisfying $\|\tilde{\boldsymbol{y}} - \bF \boldsymbol{x}\|_2 \leq \epsilon^{\prime}$ given that $\|\tilde{\boldsymbol{y}} - \bF \bG \boldsymbol{x}\|_2  = \|\boldsymbol{\eta}\|_2 \leq \epsilon$. We also arrive at appropriate values of the constants $C_0$ and $C_1$ to derive the bound stated in our Theorem \ref{thm:recov}. \QEDA
\\
Note that this minimization problem does not account for the actual matrix $\boldsymbol{F_t}$ being known or even estimated during the recovery process. In practice, using the algorithm we have proposed in section \ref{sec:alg}, the matrix $\boldsymbol{\Delta}$ (and hence $\boldsymbol{F_t}$) is estimated at each step, and the recovery is realistically, much better than the bound arrived at using the approach above. In future work, we hope to be able to also provide a bound for this scenario where $\boldsymbol{F_t}$ is estimated.

\section{Conclusions and Discussion}
\label{sec:conclusion}
We have presented a method to correct for perturbations in a compressive Fourier sensing matrix \textit{in situ} during signal reconstruction. Our method is simple to implement, robust to noise and well grounded in theory. We have discussed several applications of our framework. Moreover, we have proved conditional convergence of our algorithm to a local optimum, and shown that the basic computational problem has a unique solution under reasonable conditions. We conjecture that due to the special structure of our problem, the requirements on the number of measurements is much below what is predicted by the theoretical development so far. In the case when $P = M = N$, we prove the uniqueness of the solution to a problem that minimizes a linear approximation to the original objective function. For the main algorithm and its analysis, however, we have consciously avoided using a Taylor approximation (Baseline 2) for the algorithm presented unlike \cite{zhang2012robustly,nehorai2014structured}, even though it may initially appear to simplify the problem considerably. The primary reason for this is to avoid introduction of modeling error due to the Lagrange remainder term which can be quite significant except at small values of $r$. Our experimental results justify this choice.

Future work will involve proving analytical bounds for the global optimum of Algorithm \ref{A1}, which we believe will be stronger than those provided by results from standard CS \cite{Candes2008}, MMV \cite{Duarte2011} or GMMV \cite{Heckel2012}. We also aim to explore our algorithm in the context of different sampling strategies in practical MRI acquisition or various modes of tomographic acquisition. Furthermore, the problem of mismatch of \emph{both}, the Fourier sensing matrix and the signal representation matrix, is an interesting avenue for research. 

\bibliographystyle{IEEEtran}
\bibliography{refs}


%








\end{document}